\newcommand{\simgt}{\,\rlap{\lower 3.5 pt \hbox{$\mathchar \sim$}} \raise
1pt \hbox {$>$}\,}
\newcommand{\simlt}{\,\rlap{\lower 3.5 pt \hbox{$\mathchar \sim$}} \raise
1pt \hbox {$<$}\,}

\def\be{\begin{equation}}
\def\ee{\end{equation}}
\def\ba{\begin{eqnarray}}
\def\ea{\end{eqnarray}}

\def\lra{\leftrightarrow}

\def\dmeV{\delta m^2_{\rm eV}}
\def\siggi{\mbox{\boldmath $\sigma$}}
\def\rma{{\rm a}}
\def\rmb{{\rm b}}
\def\QCD{{\rm QCD}}
\def\BBN{{\rm BBN}}
\def\NSE{{\rm NSE}}
\def\fr{{\rm fr}}
\def\res{{\rm res}}
\def\mpl{m_{\rm pl}}
\def\Ac{A_{\rm c}}
\def\Ae{A_{\rm e}}
\def\An{A_{\rm n}}
\def\Ap{A_{\rm p}}
\def\Anua{A_{\nu_\alpha}}
\def\Anue{A_{\nu_{\rm e}}}
\def\Anus{A_{\nu_{\rm s}}}
\def\Anumu{A_{\nu_\mu}}
\def\Anutau{A_{\nu_\tau}}
\def\nB{n_{\rm B}}
\def\nn{n_{\rm n}}
\def\np{n_{\rm p}}

\def\AB{A_{\rm B}}
\def\Nnueff{N_\nu^{\rm eff}}
\def\GF{G_{\rm F}}
\def\nue{\nu_{\rm e}}
\def\nus{\nu_{\rm s}}
\def\bs{b_s}
\def\bc{b_c}
\def\TC{T_{\rm res}}
\def\st{{\rm qs}}

\def\l{\left}
\def\r{\right}
\def\ot{\frac{1}{2}}
\def\ra{\rightarrow}

\def\GeV{{\,{\rm GeV}}}
\def\MeV{{\,{\rm MeV}}}
\def\eV{{\,{\rm eV}}}

\def\VmV{\delta V}
\def\dPz{\delta P_z}
\def\pperp{{\bf P}\!_\perp}

\def\dm{\delta m^2}
\def\He{$^4$He }
\def\Hec{$^4$He}
\def\Pc{P_{\rm c}}

\def\pot{{\rm pot}}
\def\kin{{\rm kin}}
\def\tot{{\rm tot}}
\def\alfa{\vartheta}
\def\beda{\varphi}
\def\ddt{\frac{d}{dt}}

\def\ot{\frac{1}{2}}

\def\till{(P_z^- + \Pc)}
\def\alp{\alfa^+}
\def\alm{\alfa^-}
\def\bav{\beda^+}
\def\bwi{\beda^-}
\def\salp{\sin 2\alp}
\def\salm{\sin 2\alm}
\def\calp{\cos 2\alp}
\def\calm{\cos 2\alm}
\def\sbav{\sin \bav}
\def\sbwi{\sin \bwi}
\def\cbav{\cos \bav}
\def\cbwi{\cos \bwi}
\def\l{\left}
\def\r{\right}

\documentclass[12pt]{report}
\usepackage[dvips]{epsfig}

\begin{document}

\begin{titlepage}\parindent=0pt\parskip=0pt\vspace*{-11mm}

\vspace{5mm}\par
\begin{center}
  \LARGE\bf  
  Sterile Neutrinos \\in Big Bang Nucleosynthesis
  \\[25mm]
  \large 
  Diplomarbeit \\ 
  von \\ 
  Robert Buras\\[25mm]
  M\"unchen, Dezember 1999
\end{center}

\vspace{\fill}\par

\begin{center}
\begin{picture}(51,40)
\end{picture}
\\[10mm]

  \large
Physik-Department der Technischen Universit\"at M\"unchen \\
Institut f\"ur Theoretische Physik T39 \\
Prof. Dr. Wolfram Weise \\
Ausgefertigt am Max-Planck-Institut f\"ur Physik\\
(Werner-Heisenberg-Institut)\\
bei Dr. Georg Raffelt\\
\end{center}

\vspace{5mm}\par

\end{titlepage}

\pagenumbering{roman}

\centerline{\bf Zusammenfassung}\bigskip

\noindent Neutrinophysik hat in diesem Jahrzehnt sehr an Popularit\"at gewonnen, nicht zuletzt
aufgrund der Messung von Neutrinooszillationen. Allerdings lassen sich nicht gleichzeitig
alle Oszillationsexperimente vollst\"andig durch das mit Neutrinomassen erweiterte Standardmodell erkl\"aren.
Eine m\"ogliche L\"osung bietet die Einf\"uhrung eines vierten Neutrinos, das jedoch aufgrund
der Z$^0$ Resonanzbreite nicht schwach wechselwirken darf.

Ein solches steriles Neutrino k\"onnte aufgrund seiner gravitativen Wechselwirkung sowie
der Neutrinooszillationen einen Einflu\ss\ auf die primordiale Nukleosynthese haben.
Dieses Ph\"anomen ist schon in einer Vielzahl von Ver\"offentlichungen behandelt worden.
Es hat sich jedoch als sehr schwierig herausgestellt, das nicht-lineare Differentialgleichungssystem,
da\ss\ die Neutrinooszillationen im fr\"uhen Universum beschreibt, zu l\"osen.

W\"ahrend bisherige Publikationen sich meistens auf numerische Rechnungen beschr\"ankten,
untersuchen wir das System analytisch, wobei wir als einzige N\"aherung die Impulsverteilung
vernachl\"assigen.
Dabei achten wir besonders auf
das Verhalten des Systems kurz nachdem die Resonanztemperatur unterschritten worden ist. Im
Einklang mit anderen Publikationen sehen wir einen exponentiellen Anstieg in der Neutrinoasymmetrie.
Wir beweisen, da\ss\ f\"ur einen signifikanten Bereich der Mischungsparameter $\delta m^2$ und $\sin^2 2\theta$
diese Neutrinoasymmetrie zu oszillieren beginnt und sogar das Vorzeichen wechselt.
Damit werden numerische Kalkulationen best\"atigt, die diesen Effekt schon fr\"uher entdeckt haben,
deren numerische Stabilit\"at jedoch angezweifelt wurde.

\newpage

\centerline{\bf Acknowledgement}\bigskip

\noindent I would especially like to thank my supervisor Georg Raffelt, who introduced me to this
exciting subject and helped me prepare for a life as a physicist.

I would also 
like to thank Wolfram Weise for his willing support, and for teaching me quantum
mechanics.

Thanx to Klaus B\"ocker for listening to all my crazy ideas, and to
Bj\"orn P\"otter for his computer guidance.

Extra thanx to Mr. Hiermeier who never gave up showing me the fun in Mathematics.
Peace to everybody else who should be standing here, I'd never be finished.

\newpage

\pagenumbering{arabic}

\tableofcontents

\chapter{Introduction}

Neutrino physics has become very popular during the past decade.
The main impetus in this field has come from the strong improvements in detecting
neutrinos;
there is now strong evidence from two natural neutrino sources,
the solar core \cite{solarexp} and the Earth's atmosphere \cite{Atmos}, that neutrinos of a certain flavor
disappear on their way to the detector. Furthermore, a
neutrino beam produced in the laboratory
has been found to change flavor in the LSND experiment \cite{99LSND}.

The most natural explanation for these anomalies is achieved by extending
the Standard Model with neutrino masses. Together with the Cabibbo-Kobayashi-Maskawa (CKM)
matrix for the lepton sector, the masses imply flavor mixing;
neutrino oscillations have been born \cite{P57}. Such oscillations
can be fully parameterized by the differences between the squared masses $\dm_{ij}\equiv
m_j^2-m_i^2$ and
the mixing angles $\theta_{ij}$. Also, a possible phase in the CKM matrix could
induce CP violation. With these neutrino oscillations, the experiments can be explained
by the solutions presented in Table \ref{noexp}.

However, with this straightforward extension of the Standard Model
it is not possible to explain all three experiments simultaneously, simply because
the trivial condition
\be \sum \dm_{ij} = (m_3^2-m_2^2) + (m_2^2-m_1^2) + (m_1^2-m_3^2) = 0 \ee
cannot be fulfilled by the data.
Hence, many interesting new models have been suggested to explain the three different
experiments. One of the most appealing possibilities is the existence of
a fourth neutrino. Since the $Z^0$ decay width \cite{PDG98a} excludes more than three standard
neutrinos, this new flavor must be inert with respect to the standard weak
interaction.

\begin{table}[ht]
\[\begin{array}{llll}
\hline \hline
{\rm Experiment} & {\rm Favored~Channel} & |\dm| [\eV^2] & \sin^22\theta \\ \hline
{\rm Solar} &&& \\
~~~{\rm Vacuum} &\nu_{\rm e}\ra {\rm anything} & (0.5\mbox{--}8)\times 10^{-10} & 0.5\mbox{--}1 \\
~~~{\rm MSW~(small~angle)}
 &\nu_{\rm e}\ra {\rm anything} & (0.4\mbox{--}1)\times 10^{-5} & 10^{-3}\mbox{--}10^{-2} \\
~~~{\rm MSW~(large~angle)}
 &\nu_{\rm e}\ra \nu_\mu ~{\rm or}~ \nu_\tau & (3\mbox{--}30)\times 10^{-5} & 0.6\mbox{--}1 \\
{\rm Atmospheric} & \nu_\mu \ra \nu_\tau & (1\mbox{--}8)\times 10^{-3} & 0.85\mbox{--}1 \\
 & \nu_\mu \ra \nu_{\rm s} & (2\mbox{--}7)\times 10^{-3} & 0.85\mbox{--}1 \\
{\rm LSND} & \bar \nu_\mu \ra \bar \nu_{\rm e} & 0.2\mbox{--}10 & (0.2\mbox{--}3)\times 10^{-2} \\
\hline\hline
\end{array} \]
\caption{{\it Results from neutrino oscillation experiments \cite{raffi99a}.
The values are nominal $2\sigma$ ranges.}}
\label{noexp}
\end{table}

Although the hypothesis of such a sterile neutrino may seem highly speculative,
its possible existence is the most far-reaching implication of the
current experimental situation. It is certainly
worthwhile to investigate its consequences! Sterile neutrinos would have a strong
impact on certain astrophysical environments due to their mixing with active neutrinos. These
effects would become especially important under extreme conditions, like in core-collapse supernovae
\cite{ed} and in the Early Universe.

It is the impact of sterile neutrinos in the Early Universe, and in particular the
outcome of the primordial nucleosynthesis, that has been the subject of 
an intense recent debate.
By incorporating sterile neutrinos into the well understood mechanism of standard
Big Bang nucleosynthesis (BBN), two frontiers of physics would be connected in one
phenomenon, making this a powerful tool for predictions.

The production of primordial nuclei depends sensitively on the expansion
rate of the Universe, which in turn depends on the number density of sterile neutrinos.
This simple correlation is well known \cite{KTTEU}; the actual challenge is to
deduce how much the sterile neutrino sector is populated.
Since the only connection between sterile neutrinos and ordinary matter,
apart from gravity, is given by oscillations, this phenomenon must be the key for creating
a relation between sterile neutrino parameters and Big Bang nucleosynthesis predictions.

In the beginning of the 1990's, the number density for sterile neutrinos
at the time of primordial nucleosynthesis was calculated 
depending on the parameters for oscillations between an active
and a sterile neutrino, $\dm$ and $\theta$ \cite{EKMBD90f,BD91a}.
The idea was to constrain the allowed parameter space, since strong 
observational results (which have weakened since)
set an upper bound on the number density of the $\nu_{\rm s}$.
In these calculations, the influence
of the thermal plasma on the mixing parameters was included, but the small CP
asymmetric contributions were neglected. The constraints were so strong that
they clearly excluded the $\nu_\mu\leftrightarrow\nu_{\rm s}$ solution
for the atmospheric neutrino anomaly.

In 1995, Foot, Thomson and Volkas \cite{FTV96a} found an interesting effect
by including the CP asymmetric contributions; the asymmetry could
induce different population rates for the sterile neutrinos and antineutrinos,
or equivalently, different depopulation rates for the active neutrinos and
antineutrinos. This would change the neutrino asymmetry and thus would have an
influence on the population rates. They found that this back-reaction could
amplify an initially small CP asymmetry by several orders of magnitude for a
large range of parameter space.
(Actually, this back-reaction had been
discussed much earlier by Barbieri and Dolgov \cite{BD91a}, but they
erroneously found the effect to be small.)

The implications from this new effect were manifold. The most interesting conclusion
was given by a model in which $\nu_\tau\leftrightarrow\nu_{\rm s}$ oscillations created
a large $\nu_\tau$ asymmetry, which in turn suppressed $\nu_\mu\leftrightarrow\nu_{\rm s}$
oscillations [12--17]. 
In such a model,
the atmospheric neutrino anomaly could be explained by $\nu_\mu\leftrightarrow\nu_{\rm s}$
oscillations without the sterile neutrinos coming into thermal equilibrium.
Another thoroughly discussed implication treated the impact of a large
$\nu_{\rm e}$ asymmetry on primordial nucleosynthesis \cite{FV96a,FV97a,BFV98a};
the large chemical potential $\mu_{\rm e}$ would change the neutron-to-proton ratio
and thus the nuclear abundances.

Clearly, sterile neutrinos can only be included correctly into BBN
if the mechanism producing large neutrino asymmetries is revealed.
Unfortunately, this has proven to be a very difficult task.
A series of papers have been published on this subject by several
groups [12--36], 
but the situation remains unclear. The main problem is that
the system of differential equations describing the mixing is very complex: one has
to treat neutrinos of different momentum separately, and one especially has to
take care of the non-linear terms in the equations. Therefore, most works tried
to solve the problem with numerical calculations. Thereby, they often used
either the adiabatic approximation, which simplifies the equations for a given
momentum, or they neglected the momentum distribution. Only during the past
year some authors claim to have
solved the full system of differential equations numerically \cite{SF98a,F98a,FV99a}.

Due to these difficulties in solving the system,
the different works present contradictory solutions: when neglecting the
momentum distribution, the neutrino asymmetry shows an oscillating behavior \cite{S96a,EKS99a}.
The system then ends up with a
calculable value of the $\nu$ asymmetry, but with unpredictable sign. Such a scenario would introduce
domains in the Early Universe with different signs of neutrino asymmetry
\cite{SF99a}. On the other hand, applying the adiabatic approximation \cite{FV96a} is
questionable since the neutrino oscillations are not adiabatic close to the
resonance for the parameter space of interest. Here the neutrino asymmetry does
not oscillate. Since the numerical calculation using the full system of differential
equations is very CPU-time consuming, one also has to question whether
the calculations are done with sufficient accuracy and whether the numerics are
stable at all. After all, non-linear systems tend to show chaotic behavior.
Besides, different results achieved by these calculations are contradictory on the point
of oscillating neutrino asymmetry \cite{F98a,SF99a}.

All these numerical works have in common that they predict a similar final
absolute neutrino asymmetry (if effects of neutrino domains are ignored) of orders
$10^{-2}\mbox{--}1$. But even this outcome has been questioned very recently in an
analytical work by Dolgov {\it et al.} \cite{SEMIKOX}, who end up with a
final asymmetry several orders of magnitude lower than the former results.
Thus, the only point on which all works agree is that
oscillations between sterile and active neutrinos in the Early
Universe have the potential of creating a neutrino asymmetry at least of order
$10^{-5}$, a number which is still large compared to the baryon asymmetry, $\eta=\mathcal{O}(10^{-9})$.

The main problem is to get a grip on the exact behavior of the non-linear system.
Our work will try to bring some clarity into this subject. Most of the former
papers were strongly based on numerical calculations. In this paper, the
discussion will be based on a thorough analytical treatment, which will be
supported by simple numerical calculations. We will describe the evolution of
the system, neglecting the momentum distribution, and compare our results with a
numerical solution for the parameters $(\dm,\sin 2\theta_0)=(-1 \eV^2,10^{-4})$.
It should be the job of future investigations to include the
effects from the momentum distribution.

In Chapter 2, those aspects of the Early Universe and Big Bang nucleosynthesis
will be summarized which are important for our system. Chapter 3 contains all relevant
information about neutrino oscillations in matter. In Chapter 4, we analyze
the system in the simple two flavor case. After introducing initial conditions,
we prove that our system creates oscillations of the lepton asymmetry.
In Chapter 5, we summarize our findings.

\chapter{Big Bang Nucleosynthesis}

We study the influence of neutrinos, in particular their effective number of flavors and
the $\nu_{\rm e}$ chemical
potential, on Big Bang nucleosynthesis (BBN). To this end
we first describe the expansion of the Early Universe before and
during BBN. We define number density and lepton asymmetry.
Next, we give a short summary of BBN, concentrating on the
main product, \Hec, and we estimate the influence the neutrino sector can have
on its abundance. Finally, we summarize the observational results.

\section{Dynamics in the Radiation Epoch}
\label{EXPANsec}

In order to discuss the influence of neutrinos on BBN we need to introduce
some general concepts pertaining to the relevant cosmological epoch. Naturally,
we only need to look at the Universe at
times before and during BBN, i.e. at temperatures above 0.1 MeV. On the other
hand, for the mixing parameters we will consider,
neutrino oscillations are strongly damped at temperatures much higher than
the QCD phase transition scale $\Lambda_\QCD$ of around $200 \MeV$.
The temperatures of interest are therefore between $200 \MeV$ and $0.1 \MeV$,
corresponding to cosmological time scales from $10^{-5}$ s to $3$ min.

The Universe during this epoch is flat, homogeneous and isotropic, and can therefore
be described by the Robertson-Walker metric. The Friedmann equation for these
conditions is simply \cite{KTTEU}
\be H^2(t)=\frac{8\pi G}{3}\,\rho(t), \label{friedeq} \ee
where $H(t)=\dot R(t)/R(t)$ is the Hubble parameter or expansion rate, $R$
the cosmic scale factor, $G=\mpl^{-2}$ Newton's constant, and $\mpl=1.22\times10^{19} \GeV$
the Planck mass.
Because the energy density $\rho$ is dominated by radiation, it scales as
$R^{-4}$, which upon integrating (\ref{friedeq}) gives
$ H= \ot\, t^{-1}$.

The radiation density is usually expressed in the form
\be \rho = \frac{\pi^2}{30}\,g_\ast T^4 \label{rho2T}, \ee
where $T$ is the photon temperature and $g_\ast$ the total effective number of
relativistic degrees of freedom,
\be g_\ast = \sum_{i={\rm bosons}} g_i \l(\frac{T_i}{T}\r)^4 +
 \,\frac{7}{8} \sum_{i={\rm fermions}} g_i \l(\frac{T_i}{T}\r)^4. \ee
Here, $g_i$ is the number of internal degrees of freedom of particle
species $i$ and
$T_i$ is its temperature. If we insert (\ref{rho2T})
into (\ref{friedeq}) and use $t=1/2H$, we finally derive a relationship
between time and temperature:
\be t =\ot \sqrt{\frac{90}{8\pi^3}}\; g_\ast^{-1/2}\,\frac{\mpl}{T^2}
      = 0.301\, g_\ast^{-1/2}\,\frac{\mpl}{T^2}.
    \label{Hgeneral}\ee
As another consequence, we note that $T\propto R^{-1}$ as long as $g_\ast=\mbox{const}$.

In the range $1 \MeV \simlt T \simlt 100 \MeV$, the only particles which contribute
significantly to the radiation density
are photons ($g_\gamma=2$), electrons and positrons ($g_e=4$), and
three left-handed neutrino families with $g_\nu=2$ each. For all of them $T_i=T$
applies as long as they remain in thermal equilibrium so that $g_\ast = 10.75$.
Any additional neutrino species would add another $7/8$ per internal degree of
freedom. For $T \simgt 100 \MeV$, $g_\ast$ is higher due to
the presence of muons and pions, and for $T\simgt\Lambda_\QCD\approx 200 \MeV$ many
gluon and quark degrees of freedom are excited.
For $T \simlt 1 \MeV$, the electrons and positrons become non-relativistic
and annihilate, reducing $g_\ast$ by $(7/8)g_e=7/2$. This effect heats the photons (the
neutrinos are already decoupled) so that $T_\nu/T$ is reduced to $(4/11)^{1/3}$,
leading to $g_\ast=3.36$.

In our subsequent discussion we will ignore these effects and always use the
value $g_\ast = 10.75$, implying
\be H = \ot\, t^{-1} = 5.5\, \frac{T^2}{\mpl}. \label{Hspec}\ee
This is a good approximation as long as the oscillation parameters are
such that the crucial events happen in the range $1 \MeV \simlt T \simlt 100 \MeV$.

\section{Number Densities and \\ Fermion Asymmetries}

Because we want to study the influence of
neutrino number densities and matter-antimatter asymmetries,
we introduce appropriate measures for them. It is
convenient to normalize these quantities to the photon number density
\be n_\gamma = 2\,\frac{\zeta_3}{\pi^2}\,T^3, \label{photondensity} \ee
where $\zeta$ is the Riemann zeta function with $\zeta_3\approx1.202$.

Fermions may have non-vanishing chemical potentials $\mu_i$ which
affect their number densities. In addition, a non-vanishing $\mu_i$ implies a CP
asymmetry for species $i$ which we parameterize as
\be A_i\equiv \frac{n_i - \bar n_i}{n_\gamma}, \ee
where $\bar n_i$ refers to the antiparticle density.

At temperatures below the QCD phase transition
virtually no
hadronic antimatter exists. Since there are no baryon-number
violating interactions at this late epoch, the total baryon number is conserved.
Therefore, the number density of baryons, $\nB$,
scales as $R^{-3}$ or $T^3$. The baryon asymmetry
\be \AB = \frac{\nB -\bar \nB}{n_\gamma}\approx\frac{\nB}{n_\gamma} \ee
is thus constant under the cosmic expansion.

The commonly used present-time baryon asymmetry $\eta$ is related to our $\AB$ through
\be \eta = \frac{4}{11} \AB = 1 \mbox{ to } 6 \times10^{-10}.\ee
The difference is due to photon heating when electrons and positrons
annihilate, an effect which modifies all asymmetries.
We will always take $A_i$ to refer to the fermion asymmetry of species $i$ at the epoch before
${\rm e}^+{\rm e}^-$ annihilation.

For $T<\Lambda_\QCD$,
almost all baryons are either
neutrons n or protons p, which implies that
$\nB = \nn+\np$. The non-trivial
evolution of $\nn$ and $\np$ will be discussed in the next section.

We will assume that the Universe is charge neutral. Neglecting muons and pions, we
therefore use
\be\Ae=\Ap \ee
for the electron asymmetry.

Turning to neutrinos, we can safely neglect their mass. Therefore,
their number densities are given by
\be n_\nu = \int\frac{d^3{\bf p}}{(2\pi)^3}\, f(p,T,\mu), \ee
where $E=p=|{\bf p}|$, $f(p,T,\mu)=[e^{(p-\mu)/T}+1]^{-1}$ is the Fermi-Dirac phase-space distribution
function, and $\mu$ is the chemical potential. For
very small asymmetries ($\mu\ll T$) we find
\be n_\nu+ \bar n_\nu=\frac{3}{4}\, n_\gamma +{\mathcal{O}}\l((\mu/T)^2\r) \ee
and
\be A_\nu=\frac{\pi^2}{12\zeta_3}\,\frac{\mu}{T} + {\mathcal{O}}\l((\mu/T)^3\r)
   \approx 0.68\,\frac{\mu}{T} + {\mathcal{O}}\l((\mu/T)^3\r), \label{ass}\ee
where we have used that the chemical potential for antineutrinos is $\bar \mu=-\mu$.
Of course, these equations are only valid as long as the
neutrinos are in thermal equilibrium.

\section{Helium Abundance}

Big Bang nucleosynthesis is a rather complex system depending on a
number of parameters, including the baryon asymmetry $\eta$ and the effective
number of neutrino families $\Nnueff $, or more generally, $g_\ast$.
This system has been analyzed very thoroughly, including exhaustive numerical
calculations. Here we will give a short summary, concentrating on the production
of \Hec.

The outcome of BBN will be abundances of the different species $(A,Z)$ with $Z$ protons and
$A-Z$ neutrons. We will express these abundances as the mass fraction contributed
by the species, i.e.
\be X_{A,Z}\equiv\frac{A \,n_{A,Z}}{\nB}, \ee
where $n_{A,Z}$ and $\nB$ are the number densities of the nuclear species $(A,Z)$ and
all baryons, respectively.

In strict thermodynamic equilibrium, the abundances are given by
nuclear statistical equilibrium (NSE) so that
\be (X_{A,Z})_\NSE \propto \eta^{A-1} \exp\l(\frac{-B_{A,Z}}{T}\r), \label{XAeq} \ee
where $B_{A,Z}$ is the binding energy of the nuclear species $(A,Z)$. We see that the
NSE abundances are strongly suppressed by the small baryon asymmetry
$\eta \approx 10^{-9}$. This suppression
is compensated by the exponential factor in (\ref{XAeq}) at low
temperatures $T\ll B_{A,Z}$. For \Hec, $(X_{\rm He})_\NSE \approx \mathcal{O}(1)$ at
$T\approx0.3\MeV$. Heavier nuclei have significant NSE abundances only
at even lower temperatures.

The NSE abundance for an element X will freeze out when the rates \hbox{$\Gamma({\rm ab}\lra{\rm X})$} of
producing it from the lighter elements a and b become smaller
than the expansion rate $H$. We have
\be \Gamma(\mbox{ab}\ra \mbox{X})
      \propto n_\rma n_\rmb\, \exp\l[-2 \l(\frac{k}{T_\MeV}\r)^{1/3}\r], \label{Gabcd}\ee
where $k\approx Z_\rma^2 Z_\rmb^2A_\rma A_\rmb/(A_\rma + A_\rmb)$ and $T_\MeV=T/\MeV$. The first two
factors are the number densities of the nuclear species a and b, respectively,
the last factor represents the Coulomb-barrier suppression, which increases with
$A_i$ and $Z_i$. The freeze-out temperatures increase with the Coulomb-barrier
suppression.

As the temperature decreases, the
NSE abundances increase, but at the same time the nuclear reactions begin to
freeze out. Therefore, heavy nuclei are not produced during BBN because they
freeze out long before their NSE abundances have become significant.

Apart from traces of other nuclei, BBN produces primarily \He so that it is a
good approximation to assume that all neutrons end up in \Hec. Then
\be Y \equiv X_{\rm He} \approx \frac{4n_{\rm He}}{\nB}=\frac{4(\nn/2)}{\nn+\np}=
\frac{2(\nn/\np )_\BBN }{1+(\nn/\np )_\BBN }. \label{Y} \ee
Therefore, the all-important helium mass fraction $Y$ depends primarily on the
n/p ratio at the time when \He freezes out of NSE, which happens at
$t_\BBN =$ 1\mbox{--}3 minutes.

To find $(\nn/\np )_\BBN $, we have to follow the evolution of the
n/p ratio from the beginning. At temperatures above $1 \MeV$, reactions of the type
\be {\rm n}+\nue \longleftrightarrow {\rm p}+{\rm e},  \ee
maintain chemical equilibrium with the rate $\Gamma\propto\GF T^5$. Therefore
\be \l(\frac{\nn}{\np}\r)
    = \exp\l[-\frac{Q}{T}+\frac{\mu_{\rm e}}{T}-\frac{\mu_{\nue}}{T}\r], \label{npgen}\ee
where $Q=1.293 \MeV$ is the mass difference between neutrons and protons,
and $\mu_{\rm e}$ and $\mu_{\nue}$ are the chemical potentials of the electrons
and electron neutrinos, respectively. Surely, $\mu_{\rm e}$ can be neglected
since $\mu_{\rm e}/T \approx \eta \approx 10^{-9}$. For the moment we will set the neutrino
asymmetry to zero which is equivalent to $\mu_{\nue}=0$. Then the n/p ratio
depends only on the temperature.

At $T_\fr \approx 0.8 \MeV$, the n/p ratio freezes out and thus
\be \l(\frac{\nn}{\np}\r)_\fr \approx \exp(-Q/T_\fr ) \approx 0.2.\ee
More exact calculations \cite{KTTEU} give $(\nn/\np )_\fr \approx 1/6$.

After freeze-out, the n/p ratio continues to decrease slowly due to neutron decay,
\be {\rm n} \ra {\rm p} + {\rm e}^-\! + \bar \nue. \ee
Therefore, 
\be \l(\frac{\nn}{\np}\r)_\BBN \approx
     \l(\frac{\nn}{\np}\r)_\fr  \exp\l[-\ln(2)\,\frac{t_\BBN}{\tau_{\rm n}}\r]
     \approx \frac{1}{7}, \ee
where $\tau_{\rm n}=886.7\pm1.9~ {\rm s}$ is the neutron half-life.
If we insert this value into (\ref{Y}), we get
$ Y\approx 0.25$.

\section{Non-Standard Neutrinos}

We now want to derive the influence of neutrinos on the helium abundance.
According to (\ref{npgen}), a non-zero electron neutrino asymmetry changes the n/p ratio
by a factor
\be\exp(-\mu_{\nue}/T_\fr)\approx \exp(- 1.5 \Anue)\approx (1-1.5\Anue) \ee
provided that $\Anue\ll1$.
As an example, we take $\Anue =\pm 0.01$, a realistic value according to
\cite{BFV98a}. Then $(\nn/\np)$ is altered by a factor $(1\mp 0.015)$.
Of course, this is a very rough estimate, but more detailed works
\cite{FV97a,O91a} find the same order of magnitude.

A higher effective neutrino number also alters $(\nn/\np)$, since
it raises $g_\ast$, which results in a higher expansion rate $H$ and thus in a
higher freeze-out temperature. Taking $\Nnueff=4$
instead of 3 as an example, $T_\fr$ is increased by a factor of 
\be \l(\frac{g_\ast(\Nnueff=4)}{g_\ast(\Nnueff=3)}\r)^{1/6}=\l(\frac{10.75+2\cdot\frac{7}{8}}{10.75}
\r)^{1/6}=1.025, \ee
where the power $1/6$ comes from $\Gamma/H\propto T^3/\sqrt{g_\ast}$.
This changes the n/p ratio by a factor of 1.04.

We insert these changes into (\ref{Y}) and expand to get
\be Y\approx 0.25 +0.01(\Nnueff -3)-0.33 \Anue \ee
if $\Anue \ll1$ and  $|\Nnueff-3|\simlt1$. We see that
$\Nnueff =4$ or $|\Anue |=10^{-2}$ change the helium abundance by several
percent, which is within the present experimental precision.
More detailed numerical calculations \cite{KTTEU} give
\be Y \approx 0.225 + 0.025 \log(\eta/10^{-10}) + 0.012 (\Nnueff  -3),
     \label{Yformul} \ee
where we have also included the influence of the baryon asymmetry $\eta$, since
the main uncertainty in Standard BBN comes from this parameter.

We should mention that in the literature, the variable $\Nnueff$ has often
been used not only to account for the effective number of neutrinos, but
also for the influence of $\Anue$. This was done by adding a new term $\delta \Nnueff(\Anue)$
to $\Nnueff$.

\section{Observational Results}

The best measurements of the primordial
\He abundance, Y, come from observing extra-galactic regions of ionized
H. In these systems, the abundance of heavier elements, which are not created in BBN, is
very low, so we can assume that the abundances in these regions are close to
their primordial values. The present estimate is \cite{PDG98a}
\be Y = 0.238\pm0.002\pm0.005, \ee
where the two errors are the statistical and systematic errors, respectively.
The $2\sigma$ range is then estimated \cite{FO98a} to be 0.228--0.248.

Direct present-time measurements of the cosmic baryon abundance are very
uncertain due to the dark matter problem. Much better
estimates arise from
measurements of the primordial Deuterium abundance which
depends sensitively on the baryon asymmetry $\eta$.
The best measurements come from
the absorption of quasar light by high-redshift, low-metallicity 
hydrogen clouds.
Two main results have been published \cite{BT98ab,WCLFLVB97},
\be    ({\rm D}/{\rm H})_{\rm low}  = (3.4\pm0.3)\times10^{-5} \quad \mbox{and} \quad
       ({\rm D}/{\rm H})_{\rm high} = (2\pm0.5)\times 10^{-4}, \ee
which are mutually inconsistent. These two measurements give
ranges for the baryon asymmetry of
\be \eta_{\rm low}  = 4.2\mbox{--}6.3 \times 10^{-10} \quad \mbox{and} \quad
    \eta_{\rm high} = 1.2\mbox{--}2.8 \times 10^{-10}, \ee
at a nominal $2\sigma$ level.

From the measurements of D/H and $Y$, one obtains bounds on the
effective neutrino number. Olive {\it et al.} \cite{OSW99a}
derived
\be(\Nnueff-3)_{\rm low}<0.3 \quad \mbox{and} \quad(\Nnueff-3)_{\rm high}<1.8, \ee
provided that the $\nu_{\rm e}$ chemical potential can be neglected.

The low-D result today appears to be strongly favored. If it should be confirmed, sterile
neutrinos would be forbidden to come into equilibrium for negligible
$\mu_{\nu_{\rm e}}$. However, a large positive
$\Anue$ can compensate the effect of an increased $\Nnueff$, and
can thus circumvent this constraint.

\section{Summary}

We have shown that a deviation of the effective number of neutrinos of order 1, as well as
a $\nu_{\rm e}$ asymmetry exceeding about $10^{-2}$, will have a measurable effect
on the outcome of BBN.
Sterile neutrinos have the potential to change both $\Nnueff$ and $\Anue$.
If we knew how these two variables depend on the mixing parameters of the sterile neutrinos,
we could use the measured primordial element abundances to derive constraints on
the mixing parameters.
On the other hand, if the existence of sterile neutrinos was proven by future experiments such
as MiniBooNE \cite{MINIBOONE}, a detailed understanding of their impact on primordial nucleosynthesis
would be necessary to constrain the free parameters of BBN, in particular the
baryon asymmetry $\eta$.

\chapter{Neutrino Oscillations}

Neutrino oscillations play a crucial role in our considerations of the Early
Universe involving sterile neutrinos.
In this chapter, we derive the density matrix formalism for neutrino
oscillations in media between any two neutrino flavors, active or sterile.
This includes the direct
influence of matter on the vacuum neutrino oscillations as well as scattering
processes which tend to destroy the coherence of the oscillations. Our analysis
will only be applicable for
temperatures between a few$\MeV$ and $100\MeV$ and for neutrino asymmetries
$A_\nu\ll1$.

\section{Equation of Motion}
\label{numiidenii}

Neutrino oscillations occur because the basis of the neutrino weak eigenstates
$\nu_\alpha$, $\alpha={\rm e},\mu,\tau,{\rm s},\dots$, is different from the basis of the
neutrino mass eigenstates $\nu_i$, $i=1,2,3,\dots$.
In other words, a $\nu_\alpha$ that is produced in a weak-interaction process does
not propagate like a free particle, but as a superposition of neutrinos
$\nu_i$ with different masses $m_i$, respectively. Thus,
when measured, the propagated neutrino contains contributions
of weak eigenstates other than the original $\nu_\alpha$.
From the start we include the possibility that neutrinos exist beyond the
active states $\nue$, $\nu_\mu$ and $\nu_\tau$. These additional flavors
would have to be sterile with regard to the weak interaction.

The two bases are connected by a unitary transformation $\Psi_{\rm W}=U \Psi_{\rm M}$, where
\be  \Psi_{\rm W}\equiv\l(\begin{array}{c} \Psi_{\nu_{\rm e}} \\ \Psi_{\nu_\mu} \\ \Psi_{\nu_\tau} \\
     \vdots \end{array}\r)
\quad \mbox{and} \quad
     \Psi_{\rm M}\equiv\l(\begin{array}{c} \Psi_{\nu_1} \\ \Psi_{\nu_2}   \\ \Psi_{\nu_3}    \\
     \vdots \end{array}\r) \ee
are the field vectors represented in the weak and mass basis, respectively, and
$U$ is the unitary transformation matrix.

We are mainly interested in the evolution of the weak eigenstates, as these
are the particles we can produce and measure by weak-interaction processes.
Therefore, we write the equation of motion in the basis of the weak eigenstates.
The Klein-Gordon equation is
\be (\partial_t^2 - \nabla^2 + M_W^2)\Psi_W=0, \ee
where $M_{\rm W}=UM_{\rm M}U^\dagger$ and
$M_{\rm M}={\mbox{diag}}(m_1,m_2,\dots)$ is the mass matrix in the weak and
mass basis, respectively. Of course,
$M_{\rm W}^2=UM_{\rm M}U^\dagger UM_{\rm M}U^\dagger =UM_{\rm M}^2U^\dagger$ and $M_{\rm W}$ are not
diagonal in the case of mixing.

Since the Early Universe is homogeneous, we are only interested in the time evolution
of the neutrino fields. It is therefore convenient to expand them in plane
waves $\Psi_{\rm W} = \Psi_p(t) e^{i\bf p x}$. Note that usually neutrino oscillations
are considered in environments which are spatially inhomogeneous but stationary,
e.g. in experiments involving solar or atmospheric neutrinos.
In such cases, $\Psi_{\rm W}$ would be expanded in components of fixed energy,
$\Psi_{\rm W} = \Psi_E({\bf x}) e^{-i E t}$, instead of components of
fixed momentum.

Since the neutrinos are highly relativistic, $m_\nu \ll E\approx p$, we can
linearize the Klein-Gordon equation. Then we get the usual Schr\"odinger-type equation
\cite{KPNPA,RAFFI}
\be i\partial_t\Psi_p = \Omega_p\Psi_p,  \qquad \Omega_p = p + \frac{M_{\rm W}^2}{2p}.
    \label{Schroe} \ee
We see that the off-diagonal elements of $M_{\rm W}^2$ couple the fields of the different
neutrino flavors, leading to oscillations.

An equivalent equation is given by
\be i\partial_t\rho = [\rho,\Omega],
    \label{commuteeq} \ee
where $\rho_{\alpha\beta}\equiv N \Psi_\alpha \Psi_\beta^\dagger$
is the flavor density matrix, $N$ is a normalization factor,
and we have dropped the index $p$.
The advantage of the density matrix formalism is that we 
later can include effects that destroy the coherence of the neutrino oscillations.

For later convenience, we normalize the density matrix such that for a CP symmetric neutrino
species $\nu_\alpha$ in equilibrium, $\rho_{\alpha\alpha}=1$ for all momentum modes, i.e.
\be n_{\nu_\alpha}=\int\frac{d^3{\bf p}}{(2\pi)^3}\, \Psi_\alpha \Psi_\alpha^\dagger =
\int\frac{d^3{\bf p}}{(2\pi)^3}\, f_0(p,T)\, \rho_{\alpha\alpha}(p) \label{numdens}, \ee
where $f_0(p,T)=f(p,T,\mu=0)=[1+\exp(p/T)]^{-1}$. Thus, the normalization factor $N=1/f_0(p,T)$.

We have neglected the expansion of the Universe in the derivation of the equation
of motion. It could be included by adding a term $\Omega_H=iHp\partial_p$ to the
Hamiltonian. The term disappears if we expand the field vector in comoving
plane waves, $\Psi_W=\Psi_q(t)e^{iT{\bf q x}}$, where $q=p/T$, instead of plane
waves with fixed momentum. Therefore, later in Chapter 4
we will not describe the evolution of flavor
density matrices of fixed momentum $p$, but of fixed comoving momentum $q$.

\section{Medium Effects}
\label{medef}

In the Early Universe, we are confronted with a thermal medium which interacts
with the neutrinos. Therefore, we need to include the effects of these
interactions on the neutrino oscillations.
In this section, we restrict ourselves to the discussion of the refractive
effects \cite{NR88a}.

The medium contributions to the neutrino oscillations enter the
Schr\"o\-din\-ger equation (\ref{Schroe}) through the weak-potential term
$V\equiv{\mbox{diag}}(V_{\nu_\alpha},V_{\nu_\beta},\dots)$ in 
\be \Omega = p + \frac{M_{\rm W}^2}{2p} + V. \ee
For each neutrino weak eigenstate the
contributions can be split into two terms,
\be V_{\nu_\alpha}=\pm V^A_{\nu_\alpha}-V^T_{\nu_\alpha}. \label{VisAmT}\ee
Here, the plus sign is valid for neutrinos, while
the minus sign applies to antineutrinos.

The first term accounts for the fermion asymmetries. For
neutrinos of species $\alpha$ it is given by \cite{NR88a}
\be V_{\nu_\alpha}^A = \sqrt{2} \GF n_\gamma \tilde \Anua, \label{VAdef}\ee
where $\GF=1.166\times 10^{-5} \GeV^{-2}$ is Fermi's constant and $\tilde \Anua$ a weighted sum over
all fermion asymmetries $A_i$. Generally,
\ba \tilde \Anua &=& \Anua + \Anue  + \Anumu + \Anutau \nonumber\\ && {}+ A_\alpha
           - \ot(1-4\sin^2\theta_{\rm W})(\Ae+A_\mu+A_\tau) \nonumber\\
&&  {}+\ot(1-4\sin^2\theta_{\rm W})\Ap  - \ot \An, \ea
where $\theta_{\rm W}$ is the Weinberg angle.
For our assumption of a charge-neutral Universe we have $\Ae+A_\mu+A_\tau=\Ap$ and therefore
\be \tilde \Anua = \Anua + \Anue + \Anumu + \Anutau - \ot \An + A_\alpha. \label{ANUA}\ee
Since the muons and tauons are non-relativistic, their asymmetries are negligible,
so that we will use $\Ae=\Ap$, $A_\mu=0$, and $A_\tau=0$.
In our analysis, all of the asymmetries but for one type of neutrino will
remain constant. Therefore, we write
\be\tilde \Anua = 2A_{\nu_\alpha} +\Ac,\label{Adef}\ee
where all asymmetries other than $A_{\nu_\alpha}$ have been absorbed in
a constant $\Ac$.

The second term in (\ref{VisAmT})
represents the low-energy tail of the $W^\pm$ and $Z^0$ resonances. It
has the remarkable feature that it is independent of the
CP asymmetry of the background medium and has the same effect on neutrinos
and antineutrinos. Since it is of order $\GF^2$, $V^T$ can only compete with $V^A$ in the
Early Universe, where $A\ll1$. For neutrino temperatures far below the $W^\pm$ and $Z^0$-masses
it can be written in the form \cite{NR88a}
\be V_{\nu_\alpha}^T = z_{\nu_\alpha} \GF^2 p T n_\gamma. \label{VTdef}\ee
Here, the coefficients are
\ba z_{\nu_{\rm e}}&=&\frac{7\pi^3}{45\zeta_3\alpha}\sin^2\theta_{\rm W}\l(\cos^2\theta_{\rm W}+2\r) \approx 350
\nonumber\\
z_{\nu_\mu,\nu_\tau}&=&\frac{7\pi^3}{45\zeta_3\alpha}\sin^2\theta_{\rm W}\cos^2\theta_{\rm W} \approx 97,
\ea
where $\alpha=1/137$ is the fine-structure constant.

Since we will be discussing sterile neutrinos, we stress their special role in a medium.
They do not interact, so $V_{\nu_{\rm s}}=0$. 
Likewise, a possible asymmetry $\Anus$ does not contribute to $V^A_{\nu_\alpha}$ for the active
flavor.

\section{Density Matrix Formalism for \\ Two-Flavor Oscillations}
\label{dmf}

We will now treat the simplest case of neutrino oscillations between two types
of neutrinos, so $\Psi_{\rm W}=(\Psi_{\nu_\alpha}, \Psi_{\nu_\beta})$.
The unitary transformation can be written in the simple form
\be U = \l(\begin{array}[r]{cc} \cos \theta & -\sin \theta \\
                               \sin \theta &  \cos \theta \end{array}\r), \ee
where $\theta$ is the vacuum mixing angle. Note that the absolute phase
of the off-diagonal terms is arbitrary; our convention differs from some of the
literature.
In the weak basis we get
\be \Omega = \ot\, b\, I -
         \ot\l(\begin{array}{cc} c\;\!\Delta + \VmV &    s\:\!\Delta \\
                                 s\:\!\Delta        & - (c\;\!\Delta + \VmV) \end{array}\r),
   \label{HPdmf} \ee
where $b =2p +\frac{m_1^2 + m_2^2}{2p} - (V_{\nu_\alpha}+V_{\nu_\beta})$, $I$ is the $2\times2$
unit matrix, and $\VmV=V_{\nu_\alpha}-V_{\nu_\beta}$. Furthermore, we have defined
\be s\equiv \sin2\theta,  \quad
    c\equiv \cos2\theta,  \quad
    \Delta\equiv \frac{\dm}{2p} \label{nodefs}\ee
with the mass squared difference $\dm\equiv m_2^2-m_1^2$.

To find a convenient form for the equation of motion, we represent $\rho$ and $\Omega$
by the unit matrix $I$ and the
Pauli matrices $\sigma_i$, $i=1,2,3$ \cite{RAFFI}. Then
\be \rho   = \ot\, P_0\, I + \ot\, {\bf P}\:\! \siggi, \qquad
    \Omega = \ot\, b  \, I - \ot\, {\bf B}\, \siggi, \ee
where ${\bf P}$ is a so-called flavor polarization vector,
and
\be {\bf B} = \l(\begin{array}{c} s\:\!\Delta \\ 0 \\ c\;\!\Delta + \VmV \end{array}\r).
    \label{vecB0} \ee
Inserting $\rho$ and $\Omega$ into (\ref{commuteeq}) and using the commutation
relation for the Pauli matrices,
$[\ot\sigma_i,\ot\sigma_j]=i\epsilon_{ijk}\ot\sigma_k$, we derive
\be \partial_t{\bf P} = {\bf B} \times {\bf P}.
    \label{QKE} \ee
This equation
is equivalent to the precession of a spin vector in a magnetic field. 
The physical observables are now contained in $P_z=\rho_{\alpha\alpha}-\rho_{\beta\beta}$
and $P_0=\rho_{\alpha\alpha}+\rho_{\beta\beta}$.
$f_0(p,T)P_0$ is the total occupation number of both $\nu_\alpha$ and $\nu_\beta$
of momentum mode p, while $P_z$ represents the difference between the two neutrino
types; if e.g. $P_z=1$ for all modes, then $n_{\nu_\beta}=0$. However, if $P_z=0$
for all modes, $n_{\nu_\alpha}=n_{\nu_\beta}$. The other two components $P_{x,y}$ represent
the phase of the oscillations.

\section{Scattering Processes}

We must still take into account
that the weak neutrino eigenstates scatter on the background plasma
with the rate \cite{MT94a}
\be \Gamma_{\nu_\alpha} = y_{\nu_\alpha} \GF^2 p T^4, \label{Gammadef}\ee
where $y_{\nu_{\rm e}} \approx 1.13$, $y_{\nu_\mu,\nu_\tau} \approx 0.79$ and $y_{\nu_{\rm s}} \equiv 0$.
To be precise, $y_{\nu_\alpha}$ also depends on the momentum due to the Pauli blocking
factors. We will use these momentum averaged values \cite{SEMIKOX}.

Scattering keeps the active neutrinos in thermal and chemical
equilibrium at temperatures above a few MeV.
Therefore, the integrand in (\ref{numdens}),
$f_0(p,T) \rho_{\alpha\alpha}(p)$, should be equal to $f(p,T,\mu)$.
However, $\rho_{\alpha\alpha}(p)$ changes due to the neutrino oscillations.
The scattering processes compensate this change by re- or depopulating the
number densities through $\dot\rho_{\alpha\alpha}(p)=R_{\nu_\alpha}(p)$, where the rate
is approximately given by \cite{MT94a}
\be R_{\nu_\alpha}(p) \approx \Gamma_{\nu_\alpha} \l[\frac{f(p,T,\mu)}{f_0(p,T)}-\rho_{\alpha\alpha}(p)\r]. 
   \label{Rdef} \ee
The approximation holds as long as the deviation from equilibrium is small.
Of course, $R_{\nu_{\rm s}}=\Gamma_{\nu_{\rm s}}=0$.

A much more interesting effect appears
if the scattering rates are different for the two neutrino types considered.
Then scattering distinguishes between the two $\nu$ types and thus
the coherence of the neutrino oscillations is destroyed.
Therefore, we need to include a term $-D\pperp(p)$, where
$\pperp=(P_x,P_y,0)$ and $D$ is called the damping rate.
In the case of an active neutrino oscillating with a sterile neutrino
($\beta=s$) we have $D = \ot\, \Gamma_{\nu_\alpha}$. The lengthy derivation
of this term can be read in \cite{STOD87}.

In summary, for each neutrino momentum mode $p$ we have derived a system of differential equations
\ba \partial_t{\bf P} &=& {\bf B} \times {\bf P} -D\pperp + (R_{\nu_\alpha}
-R_{\nu_\beta}){\bf \hat z}, \label{QKEFULL1}\\
    \partial_t   P_0 &=& (R_{\nu_\alpha}+R_{\nu_\beta})
    \label{QKEFULL2},\ea
\ba \partial_t{\bf \bar P} &=& {\bf \bar B} \times {\bf \bar P} -D\bar\pperp + (\bar R_{\nu_\alpha}
-\bar R_{\nu_\beta}){\bf \hat z}, \label{QKEFULL3}\\
    \partial_t \bar P_0 &=& (\bar R_{\nu_\alpha}+\bar R_{\nu_\beta}),
    \label{QKEFULL4}\ea
where ${\bf \hat z} =(0,0,1)$. In the next chapter we will study the highly non-trivial behavior of this
equation of motion.

\chapter{Oscillating Lepton Asymmetry}

We analyze the system of flavor
oscillations between active and sterile neutrinos before BBN in a simplified model.
First we discuss our approximations and present a typical numerical solution of the
simplified model. Next, we introduce a more convenient coordinate system for the analytical
treatment. With its help, we describe the evolution of our system from very
early times up to the resonance.
At resonance, we show that we can describe the evolution of the neutrino asymmetry
with a simple differential equation, and that the solution will indeed oscillate
for some of the parameter space.

\section{Simplified Model}

We describe the oscillations between two flavors of neutrinos, one active
and one sterile, in an expanding medium. This means that we neglect all
other neutrino mixings, which is a good approximation if all other effective
mixing angles are small.
Thus, we can describe our oscillations by the system of differential equations
(\ref{QKEFULL1})--(\ref{QKEFULL4}) derived in the previous chapter.
We restrict ourselves to a simplified model which we present in this
section. For definiteness, we will analyze the case where the active
neutrino is $\nu_\tau$; the analysis is
analogous for $\nu_{\rm e}$ and $\nu_\mu$.

Our most important simplification is that we neglect the momentum distribution.
This is surely not a good approximation, since then all neutrinos encounter 
the oscillation resonance at the same time. In reality only a small
fraction of the neutrinos will be simultaneously at resonance,
especially when the resonance width is small, i.e.~the vacuum mixing angle is small. 
However, the complete system is very complicated, as can be seen from the controversial literature
on its solution, and
we therefore have decided to analyze this simplified model. Of course, it
is desirable to include the effects of the momentum distribution in future
investigations.

So from now on, all neutrinos are taken to have the same momentum. We choose this momentum to
be the average of a fermion species with vanishing chemical potential,
\be \l<p\r>=\frac{7\zeta_4}{2\zeta_3} T \approx3.15\,T, \ee
where $\zeta$ is the Riemann zeta function with $\zeta_3\approx1.202$ and $\zeta_4=\pi^4/90$.
As a consequence, we only have two density matrices, $\rho_{\alpha\alpha}$ for the neutrinos
and $\bar \rho_{\alpha\alpha}$ for the antineutrinos. We normalize these in analogy to our previous
normalization in Section \ref{numiidenii}, i.e.~such that
the number densities are $n_{\nu_\tau}=n_\nu^{\rm eq}\rho_{\tau\tau}$, etc. where
$n_\nu^{\rm eq}=\frac{3}{8}n_\gamma$ is the equilibrium neutrino number density for vanishing
chemical potential. 

We will also neglect the repopulation terms $R_{\nu_\tau}$ and $\bar R_{\nu_\tau}$ defined in
(\ref{Rdef}). They will be small
if the $\nu_\tau$ and $\bar \nu_\tau$ are not depopulated significantly
by neutrino oscillations, and if $A_{\nu_\tau}\ll 1$ so that the chemical potential
has no significant influence on the equilibrium number density.
This simplification is valid for some part of the parameter space, which we will
determine later.
As a result of this approximation, $P_0$ and $\bar P_0$ are constant in time.

So now, the system of differential equations has been simplified to
\ba \partial_t {\bf      P} &=& {\bf      B} \times {\bf      P} - D
     {\bf      P}_\perp, \nonumber \\
    \partial_t {\bf \bar P} &=& {\bf \bar B} \times {\bf \bar P} - D
     {\bf \bar P}_\perp.
 \label{deq} \ea
The tau neutrino asymmetry
\be A_{\nu_\tau} = \frac{n_{\nu_\tau}^{\rm eq}}{n_\gamma}(\rho_{\tau\tau}-\bar\rho_{\tau\tau})
    = \frac{3}{8}\times\ot(P_z - \bar P_z + P_0-\bar P_0)\ee
is an important variable since it depends on $\bf P$,
enters into $\bf B$, and thus causes the system to be nonlinear.
It will be more convenient to use the equivalent variable
\be \dPz \equiv \frac{4}{3}\tilde A_{\nu_\tau} = \ot (P_z-\bar P_z) + \Pc,
   \label{dPz} \ee
where $\tilde A_{\nu_\tau}$ was given in (\ref{ANUA}) and
$\Pc \equiv \frac{4}{3}(\Anue + \Anumu - \ot \An) + \ot(P_0-\bar P_0)$ is constant.

In terms of this variable,
the coefficients in the differential equations are given by
\ba {\bf      B}(T,\dPz)&=&\l(\begin{array}{c} -\bs \\ 0 \\ -\bc + b_T - b_A \dPz \end{array}\r), \\
    {\bf \bar B}(T,\dPz)&=&\l(\begin{array}{c} -\bs \\ 0 \\ -\bc + b_T + b_A \dPz \end{array}\r),
 \ea
where we use
\ba \bs&\equiv& -~\frac{s~\dm}{2\l<p\r>}\approx-~\frac{s~\dm}{6.3}\,T^{-1}, \nonumber\\
    \bc&\equiv& -~\frac{c~\dm}{2\l<p\r>}\approx-~\frac{c~\dm}{6.3}\,T^{-1}, \nonumber\\
     b_A&\equiv& V^A_{\nu_\tau}/\dPz=k_A T^3, \nonumber\\
     b_T&\equiv& V^T_{\nu_\tau}=k_T T^5. \ea
Recall that $s=\sin2\theta$, $c=\cos2\theta$, and $\dm=m_{\nu_{\rm s}}^2-m_{\nu_\tau}^2$.
The constants are
\ba     k_A&=& \frac{3\zeta_3}{4\pi^2}\sqrt{2}\GF \approx 1.51\times 10^{-24} \eV^{-2}, 
\nonumber\\
k_T&=& \frac{14\pi}{45\alpha}\sin^2\theta_{\rm W}\cos^2\theta_{\rm W}\frac{\l<p\r>}{T}
\GF^2\,\approx 1.02 \times 10^{-44} \eV^{-4}, \ea
where $\alpha=1/137$ is the fine-structure constant and $\theta_{\rm W}$ the Weinberg angle.
The coefficient $D$ is proportional to $b_T$, so we will often use the relation
\be D=k_D b_T, \ee
where
\be k_D=\frac{\pi^2y_{\nu_\tau}}{2\zeta_3z_{\nu_\tau}}\approx 1/60, \ee
and we have used
$y_{\nu_\tau}\approx0.79$ and $z_{\nu_\tau}\approx97$ from Section \ref{medef}.
For $\nu_\mu$, the constants are the same. For $\nu_{\rm e}$, we have
$k_T\approx 3.66 \times 10^{-44} \eV^{-4}$ and $k_D\approx 1/151$.

We will be considering neutrino oscillations with small vacuum mixing angles, i.e.~$s\ll c\approx1$.
As a consequence,
the first component of ${\bf B}$ will be much smaller than the third component for most of
the time. The system becomes interesting when the third component disappears. Then
the neutrino oscillations are at resonance, i.e.~the neutrinos mix maximally.
If we assume the initial asymmetry to be negligible, the resonance condition is given by
$\bc=b_T$. This condition has a solution only if $\dm<0$. Then resonance occurs when
\be \TC=\l(\frac{|\dm| c}{6.3 k_T}\r)^{1/6}\approx 15.8 \MeV |\dmeV|^{1/6}, \ee
where $\dmeV=\dm/\eV^2$.
The resonance will be the crucial feature of the system. We will see that shortly after the
resonance, the neutrino oscillations will create a large asymmetry, an effect which
is driven by the non-linear term $b_A \dPz$ in the system of differential equations. We will
therefore only consider $\dm<0$, i.e.~the sterile neutrino is lighter than the tau neutrino.

\section{Numerical Solution}
\label{numsolsec}

\begin{figure}
\unitlength1mm
\begin{picture}(121,80)
  \put(15,0){\psfig{file=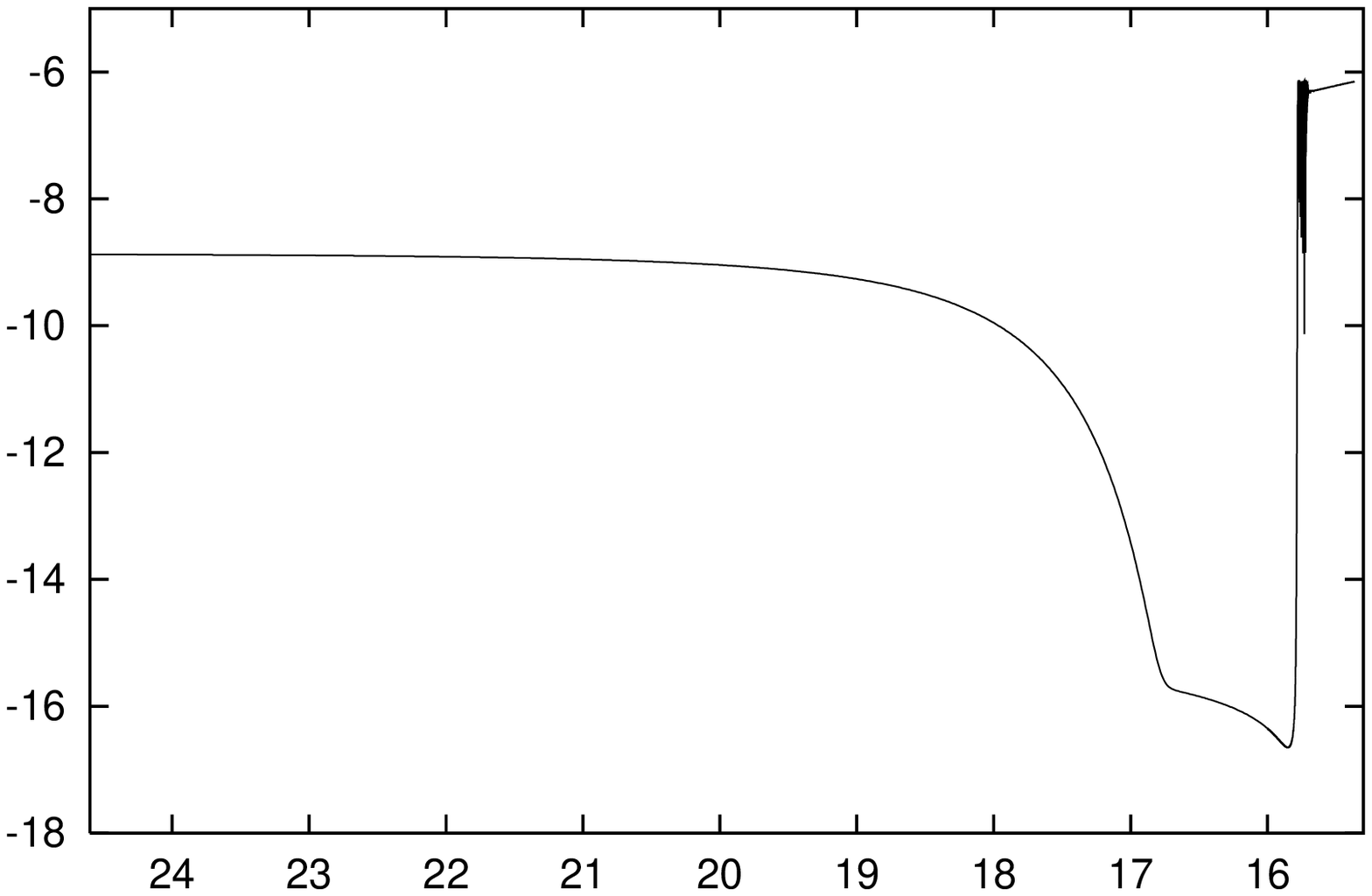,width=10.4cm}} 
  \put(115,0){\footnotesize $T [\MeV]$}
  \put(35,47){1.}
  \put(87,44){2.}
  \put(103,16){3.}
  \put(105,35){4.}
  \put(14,68){\footnotesize $\log_{10}(|\dPz|)$}
\end{picture}
  \caption{\label{dPz-evo}Evolution of $\dPz$, steps 1.~to~4.}
\unitlength1mm
\begin{picture}(121,80)
  \put(15,0){\psfig{file=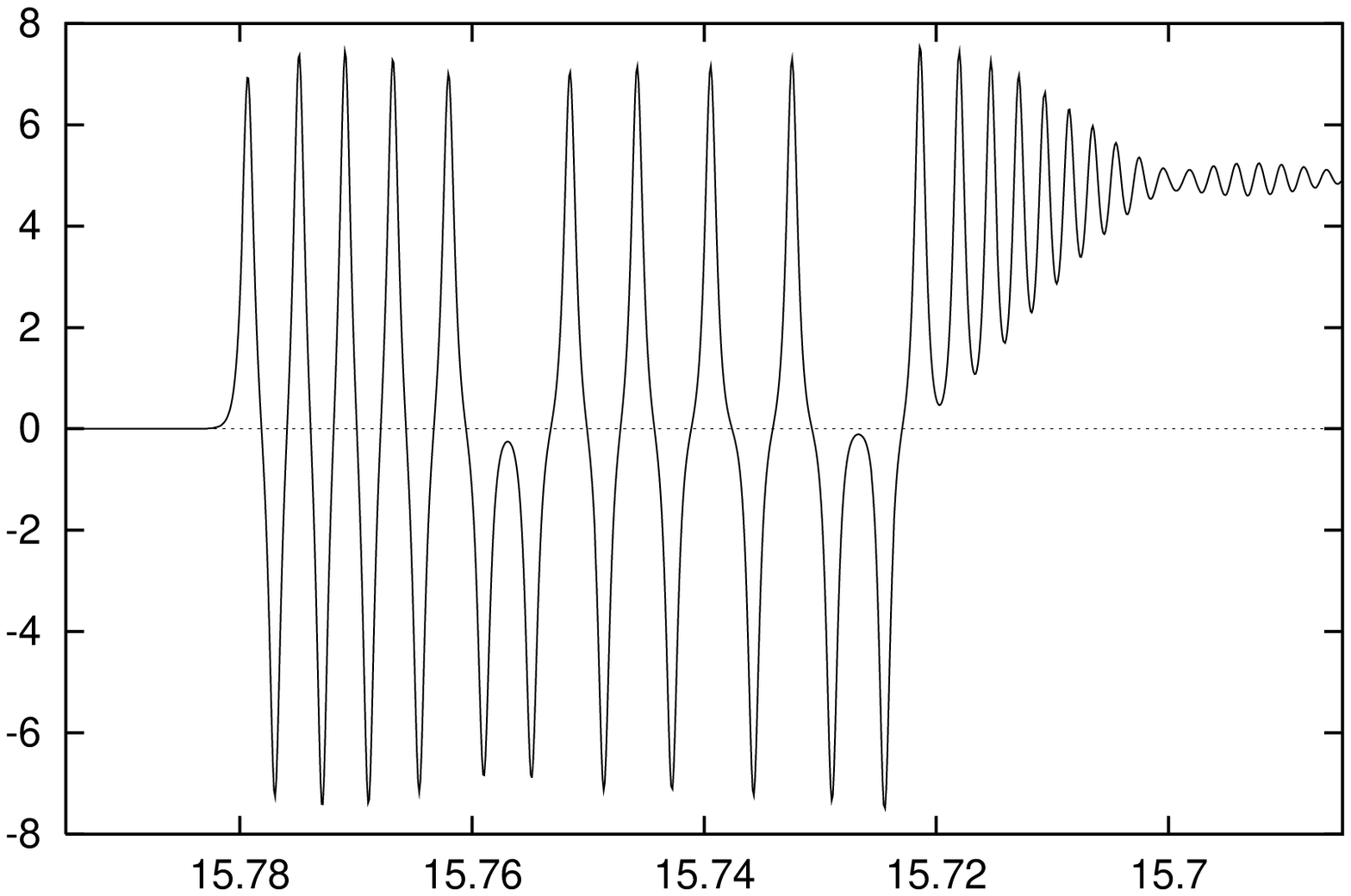,width=10.4cm}} 
  \put(115,0){\footnotesize $T [\MeV]$}
  \put(14,68){\footnotesize $\dPz [10^{-7}]$}
\end{picture}
  \caption{\label{dPz-evo-log}Oscillatory behavior of $\dPz$ during step 5.}
\end{figure}

We have solved the system (\ref{deq}) of differential equations numerically and find
results similar to those in \cite{S96a} and \cite{EKS99a}. We have plotted the
evolution of the effective asymmetry $\dPz$ for $(\dm,s)=(-1\eV^2,10^{-4})$ in Fig.~\ref{dPz-evo}.
Its behavior is representative for a large region of parameter space.
The evolution of the system falls into five distinct phases.

\begin{itemize}
\item[1.] At high temperatures, the asymmetry $\dPz$ is constant, since
damping totally destroys the coherence of the oscillations. Thus, no
flavor transition occurs at all.

\item[2.] As the temperature decreases, the mixing angle increases, while the
damping rate decreases. As a consequence, there will be a small amount of
flavor transition induced by the damping.
Due to the asymmetric term $b_A\dPz$, this transition rate will be different for 
neutrinos and antineutrinos, washing out the asymmetry $\dPz$.
However, there remains a small relic asymmetry.

\item[3.] Until the system reaches the resonance, the relic asymmetry changes slowly.

\item[4.] When the system has passed the resonance, the difference in the
transition rates for neutrinos and antineutrinos has the opposite effect than
before. In other words, a small asymmetry is no longer washed out, but amplified,
which leads to an exponential increase of $\dPz$.

\item[5.] When $\dPz=\mathcal{O}(10^{-6})$, it starts oscillating
and thereby even changes its sign. For our numerical example, we have plotted
the evolution of $\dPz$ during this step in Fig.~\ref{dPz-evo-log}.

\item[6.] Eventually, the neutrino asymmetry stops oscillating. After that, it slowly
increases.

\end{itemize}

In the past, the underlying mechanism for the oscillatory behavior in step 5 was not fully understood.
Merely the authors of \cite{S96a} discussed this phenomenon qualitatively.
Our main achievement will be to analyze the system analytically
and to prove that it indeed oscillates. Thus, we invalidate the frequently used argument that the oscillations
in this phase are an artifact caused by a numerical instability. 

\section{Spherical Coordinates}

We will now change to a more convenient coordinate system.
The two phase-related coordinates $P_x$ and $P_y$
are not very practical, so we prefer to use spherical coordinates.
As a measure for the length $P=|{\bf P}|$ of the polarization vector we use $P_z$ because it
enters directly into (\ref{dPz}). This choice becomes problematic when
$P_z=0,~P\neq0$, but as mentioned above, $P_z$ will not change significantly.
The new coordinate system consists of $P_z$ and the angles $\alfa$ and $\beda$.
These new variables are related to the old ones as
\ba P_x & = & P_z \tan\alfa\sin\beda, \nonumber\\
    P_y & = & P_z \tan\alfa\cos\beda, \nonumber\\
    P_z & = & P_z.\ea
$P_z$ represents the relation between the number densities of the active and sterile neutrinos,
$\alfa$ is a measure of how much the polarization vector deviates from the $z$-axis, and
$\beda$ is the rotation angle around the $z$-axis.
Then the differential equations (\ref{deq}) change to
\ba \dot P_z    & = & - \bs P_z \tan\alfa \cos\beda, \nonumber\\
    \dot \alfa  & = & - D \sin\alfa \cos\alfa + \bs \cos\beda, \nonumber\\
    \dot \beda  & = & \bc - b_T + b_A\dPz - \bs \cot\alfa \sin\beda, \label{sphdeq}\ea and
\ba \dot{\bar P_z}    & = & - \bs \bar P_z \tan\bar\alfa \cos\bar\beda, \nonumber\\
    \dot{\bar \alfa}  & = & - D \sin\bar\alfa \cos\bar\alfa + \bs \cos\bar\beda, \nonumber\\
    \dot{\bar \beda}  & = & \bc - b_T - b_A\dPz - \bs \cot\bar\alfa \sin\bar\beda,
    \label{sphdeq2} \ea
where over-barred quantities, as usual, refer to antineutrinos.
With (\ref{dPz}) we have a closed set of differential equations.

The only difference between the equations for neutrinos and anti\-neu\-tri\-nos is the sign of the term
$b_A\dPz$. Therefore, if this term is negligible, the variables for neutrinos and antineutrinos
develop equally.
To be more sensitive to the small differences induced by the asymmetry,
we perform another substitution and use the variables
\ba P_z^\pm   & \equiv & {\textstyle\ot} ( P_z  \pm \bar P_z ), \nonumber\\
    \alfa^\pm & \equiv & {\textstyle\ot} (\alfa \pm \bar\alfa), \nonumber\\
    \beda^\pm & \equiv & {\textstyle\ot} (\beda \pm \bar\beda) \ea
instead.
Thus $P_z^+\gg P_z^-$, $\alp\gg\alm$ and $\bav\gg\bwi$ when the fermion asymmetry is small.
Note that (\ref{dPz}) becomes
\be    \dPz(t) = P_z^-(t) + \Pc  \ee
so that $\frac{d}{dt}\dPz=\dot P_z^-$. We will make use of this simple relation by considering
$\dPz$ instead of $P_z^-$ when it is convenient.

In terms of the new variables, our system of differential equations is
\ba \dot P_z^+ & = & -\bs P_z^+ \frac{~~\salp\cbav\cbwi - \salm\sbav\sbwi}{\calp+\calm} \nonumber\\
               &   & -\bs P_z^- \frac{ -\salp\sbav\sbwi + \salm\cbav\cbwi}{\calp+\calm}, \\
    \dot P_z^- & = & -\bs P_z^+ \frac{ -\salp\sbav\sbwi + \salm\cbav\cbwi}{\calp+\calm} \nonumber\\
               &   & -\bs P_z^- \frac{~~\salp\cbav\cbwi - \salm\sbav\sbwi}{\calp+\calm}, \\
 \dot \alp &=& -\ot D \salp\calm + \bs\cbav\cbwi, \\
 \dot \alm &=& -\ot D \calp\salm - \bs\sbav\sbwi, \\
 \dot \bav &=&  \bc - b_T   \nonumber \\
&&{}+ \bs \frac{-\salp\sbav\cbwi + \salm\cbav\sbwi}{\calm-\calp}, \\
 \dot \bwi &=&  b_A \till \nonumber\\
&&{} + \bs \frac{-\salp\cbav\sbwi + \salm\sbav\cbwi}{\calm-\calp}. \nonumber\\ \ea

As a last approximation, we assume $\alfa^\pm\ll 1$, which effectively means that the polarization
vectors will stay close to the $z$-axis. We expand $\sin2\alfa^\pm$
and $\cos2\alfa^\pm$ and only take into account the leading order terms. Then we get
\ba \dot P_z^+ &= & - \bs P_z^+\l(~\alp\cbav\cbwi - \alm\sbav\sbwi\r) \nonumber\\
               &  & - \bs P_z^-\l( -\alp\sbav\sbwi + \alm\cbav\cbwi\r), \label{pzp2}\\
    \dot P_z^- &= & - \bs P_z^+\l( -\alp\sbav\sbwi + \alm\cbav\cbwi\r) \nonumber\\
               &  & - \bs P_z^-\l(~~\alp\cbav\cbwi - \alm\sbav\sbwi\r), \label{pzm2}\\
 \dot \alp &=& -D \alp + \bs\cbav\cbwi, \label{alp2}\\
 \dot \alm &=& -D \alm - \bs\sbav\sbwi, \label{alm2}\\
 \dot\bav &=&  \bc - b_T \nonumber\\
&&{} + \frac{\bs}{\alp} \l(-\sbav\cbwi +\frac{\alm}{\alp}\cbav\sbwi\r),
    \label{bav2}\\
 \dot\bwi &=&  b_A \till \nonumber\\
&&{}  + \frac{\bs}{\alp} \l(-\cbav\sbwi +\frac{\alm}{\alp}\sbav\cbwi\r).
    \label{bwi2} \ea
From (\ref{alp2}) we see that $\dot\alp\le-D\alp+\bs$. Thus $\alp\ge\bs/D$ would automatically
mean that $\dot\alp\le0$. Therefore, we can safely say that $\alp<\bs/D$ at all
times. We will see that $\alp$ becomes maximal at resonance,
so $\alp_{\rm max}\approx(\bs/D)_{\rm res}=s/(ck_D)$, where we have used $\bs/s=\bc/c$, $(\bc)_\res=
(b_T)_\res$ and $D=k_D b_T$.
Thus our approximation is only valid if $s\ll k_D=1/60$, $c\approx 1$, i.e.~for small mixing angles.

\section{Initial Conditions}

We need to derive the initial conditions before we can consider the evolution of the system of
differential equations. We begin with temperatures far above the resonance.
For $T\ra\infty$, the coefficients $D,~b_T,~R_{\nu_\tau}\propto T^5\ra \infty$, $b_A \propto T^3 \ra\infty$,
while $\bs,~\bc\propto T^{-1}\ra 0$. If we compare these coefficients with $H\propto T^2$,
we conclude that $\bs$ and $\bc$ can be neglected at very high temperatures.
Thus, from (\ref{pzp2})--(\ref{alm2}) we see that regardless of the initial conditions, $\alp$ and $\alm$
are both exponentially damped to zero, while $P_z^\pm\propto \bs$ do not change.
However, the scattering processes equilibrate $\nu_\tau$ and $\bar \nu_\tau$,
implying $\rho_{\tau\tau}$ and $\bar\rho_{\tau\tau}\ra1$. 
Since the sterile neutrinos do not interact,
$\rho_{\rm ss}$ and $\bar\rho_{\rm ss}$ remain constant. They are only diluted
whenever massive particles become non-relativistic and annihilate into the still relativistic
particles, which heats up the plasma.
For example, when the temperature decreases from $T\sim {\mbox{TeV}}$
(just before the electro-weak symmetry breaking) to $T\sim {\mbox{MeV}}$
(shortly after the quark-hadron phase transition), $\rho_{\rm ss}$ and $\bar \rho_{\rm ss}$ are
diluted by a factor of
$[g_\ast(T\sim 300 \MeV)/g_\ast(T\sim {\mbox{TeV}})]^{4/3}\sim0.07$. So if the $\nus$ were in
equilibrium with the thermal plasma at very early times by some unknown mechanism, they would have
been strongly diluted by the time they become interesting for us. We will simply assume that their
initial density is zero. Then the initial conditions are $P_0=\bar P_0=1$ and $P_z=\bar P_z=1$.
Including the small initial neutrino asymmetries, $\Anutau$ and $\Anus$,
we find
\ba P_z^+&=&1, \nonumber \\
 P_z^-&=&\frac{4}{3}(\Anutau-\Anus), \nonumber\\
\ot(P_0-\bar P_0)&=&\frac{4}{3}(\Anutau+\Anus)={\rm const}, \nonumber \\
\Pc&=&\frac{4}{3}(\Ac+\Anutau+\Anus)={\rm const}, \ea
where in $P_z^+$ we have neglected $\Anutau$ and $\Anus\ll1$. 

\section{Quasi-Static Solutions}

We will now follow the time evolution of the system. 
We first stress the important role that damping plays in our system.
We consider a variable $x$, which follows the differential equation
\be \frac{\partial}{\partial t} x(t) = -d(t)x(t)+f(t), \ee
where $d(t)$ is some damping coefficient and $f(t)$ is a function.
If $d$ and $f$ are constant, $x$ will relax to a static value
$x_{\rm st}=\frac{f}{d}$. In our case, $d$ and $f$ will slowly vary
in time, and so will the static value. But if its rate of change
$r_x\equiv\dot x_{\rm st}/x_{\rm st}$ is smaller than the damping coefficient
$d$, the damping will force the variable to follow its static value.
We will make use of this approximation; since it is not static in the strict
sense of the meaning, we will call these solutions quasi-static.

We have calculated the rates of change of the quasi-static solutions for our variables in 
Appendix \ref{AppSTAT}. For the discussion in this section it is sufficient to know that
$r=\mathcal{O}(H)$.

We can now determine the variables for large temperatures.
We can easily assume that $\bs \alfa^\pm\ll D\alfa^\pm,\bs,(\bc-b_T),\bs/\alfa^\pm$, so before
$P_z^\pm$ change significantly at all, the other four variables will have relaxed to their quasi-static
values. If we assume that $\dot P_z^-=0$, the differential equations for neutrinos and antineutrinos
(\ref{deq}) decouple. For small $\alfa$, the angular equations in (\ref{sphdeq}) become
\ba \dot \alfa  & = & - D \alfa + \bs \cos\beda \label{alfadeq}\\
    \dot \beda  & = & (\bc - b_T + b_A\dPz) - \frac{\bs}{\alfa} \sin\beda. \label{bedadeq}\ea
Since $\alfa>0$ by definition, the first term in the first equation and the second term in the second
equation are both damping terms. The two variables relax to $\dot\alfa\approx\dot\beda
\approx0$ if the quasi-static conditions are fulfilled,
i.e.~$D>H$ and $\bs/\alfa>H$, respectively. The first condition is
fulfilled for $T>{\rm few}\MeV$. When $\alfa$ has relaxed, we get from (\ref{alfadeq}) that
$\alfa\approx b_s \cos\beda/D<\bs/D$, which fulfills the second condition, $\bs/\alfa>D>H$.
Setting $\dot\alfa=\dot\beda=0$, we get the quasi-static solutions
\ba \alfa_\st&=&\frac{\bs}{\sqrt{D^2+(\bc - b_T + b_A\dPz)^2}}, \\
 \beda_\st&=&\arctan\l(\frac{\bc - b_T + b_A\dPz}{D}\r). \ea
The same is valid for the antineutrinos, but with $\dPz\ra-\dPz$.
Since $b_A\dPz\ll \bc-b_T$, we can already say that the variables
$\alm$ and $\bwi$ will be much smaller than $\alp$ and $\bav$, respectively.
Thus,
\ba \alp_\st &=&{\textstyle\ot}(\alfa_\st+\bar\alfa_\st)\approx
 \frac{\bs}{\sqrt{D^2+(\bc - b_T)^2}}, \label{alp-st}\\
 \bav_\st &=&{\textstyle\ot}(\beda_\st+\bar\beda_\st)\approx
 \arctan\l(\frac{\bc - b_T}{D}\r). \ea
The same quasi-static solutions are also derived from (\ref{alp2}) and (\ref{bav2}) if we neglect
the last term in latter.

Now it is also easy to calculate $\alm$ and $\bwi$ for a given $\dPz$ by setting $\dot \alm=0$
in (\ref{alm2}) and $\dot \bwi = 0$ in (\ref{bwi2}),
inserting $\alp_\st$ and $\bav_\st$ and setting $\cos \bwi=1$. Then
\ba \bwi_\st &=& \arcsin\l(\frac{Db_A}{D^2 +(\bc - b_T)^2}\dPz\r) \label{bwi-st} \\
 \alm_\st &=& \frac{-\bs(\bc-b_T)b_A}{[D^2 +(\bc - b_T)^2]^{3/2}}\dPz.
    \label{alm-st} \ea
Again, we need to show that the quasi-static conditions are fulfilled, which we again find to
be $D>H$.

If we insert all these variables into $\dot P_z^-$, and neglect the fourth term in
(\ref{pzm2}), we obtain
\be \ddt\dPz=\dot P_z^- = -\kappa(\dPz+\varepsilon P_z^-), \label{dpzsim}\ee
where the damping coefficient is
\be \kappa\equiv\l|\frac{2P_z^+\bs^2(\bc-b_T)Db_A}{[D^2 +(\bc - b_T)^2]^2}\r|\propto T^{-9}, \ee
and
\be \varepsilon \equiv \l|\frac{D^2 +(\bc-b_T)^2}{2P_z^+b_A(\bc-b_T)}\r|, \ee
where $\varepsilon\ll1$ for $T<\GeV$ and $T\neq\TC$. Analogously, we get
\be\dot P_z^+ = -\kappa\varepsilon P_z^+\propto T^{-7}, \ee
where we have only taken into account the first term in (\ref{pzp2}), since
the other terms are all $\mathcal{O}(P_z^-)$ smaller.
This confirms that $P_z^-$ and $P_z^+$ will not relax to their quasi-static values
at high temperatures.

\section{Evolution Towards the Resonance}

When $\kappa\simgt H$, $\dPz$ starts changing.
The damping coefficients for $\bwi$ and $\alm$ (which are $\simgt D$) are much larger than H, so
even though $\dPz$ changes, (\ref{bwi-st}) and (\ref{alm-st}) remain valid.

From (\ref{dpzsim}) we see that $\dPz$ is damped towards 0, i.e.~$P_z^- \ra -P_c$.
When $|\dPz|\ll |P_z^-|$, the second term in (\ref{dpzsim}) becomes important. We set $\dot P_z^-=0$
to get
\be (P_z^-)_\st= -\Pc~ \frac{1}{1+\varepsilon},
 \qquad (\dPz)_\st= \Pc~ \frac{\varepsilon}{1+\varepsilon}. \ee
We have to prove that $\dPz$ takes on its quasi-static value before the system passes the resonance.
The solution of (\ref{dpzsim}) is given by
\be \dPz(t)= \dPz(0) \exp\l(-\int\limits_0^t \kappa(t)dt\r). \ee
If we substitute $x=T(t)/\TC$, we can write the integral as
\be F_-(t) \equiv \int\limits_0^t\kappa\,dt =
\frac{k_D \mpl 2P_z^+k_A}{6.3^{1/6}~5.5~k_T^{1/6}}~\frac{s^2 |\dm|^{1/6}}{c^{11/6}}~ I_-(t), \ee
where
\be I_-(t_{\rm res}) = \int\limits_1^\infty \frac{x^6(x^6-1)}{[k_D^2x^{12}+(x^6-1)^2]^2}~dx\approx 293. \ee
Assuming that initially $\dPz$ and $P_z^-$ are of the same order of magnitude, we know that
$\dPz$ will decrease by a factor of order $\varepsilon\sim b_T/b_A \sim \mathcal{O}(10^{-6})$,
which corresponds to $F_-(t_{\rm res})\ge 6\ln(10)$. Thus, we obtain the condition
\be s^2 |\dm|^{1/6} \simgt 10^{-12} \eV^{1/3} \label{dpzqs}\ee
using $P_z^+\approx1$.

For $P_z^+$ we can perform a similar calculation. However, here
we demand that $P_z^+$ does {\it not} change significantly to prevent
that the repopulation terms become important.
Therefore we demand that $F_+(t_{\rm res})\ll1$, where 
\be P_z^+(t)= P_z^+(0) \exp\l[-F_+(t)\r]. \ee
Thus, 
\be F_+(t)=\int\limits_0^t \kappa(t)\varepsilon(t) dt =
   \frac{k_D \mpl k_T^{1/2}}{6.3^{1/2}~5.5}~\frac{s^2 |\dm|^{1/2}}{c^{3/2}}~ I_+(t), \ee
where
\be I_+(t_{\rm res}) = \int\limits_1^\infty \frac{x^2}{k_D^2x^{12}+(x^6-1)^2}~dx\approx15.2. \ee
We then get the condition
\be s^4 |\dmeV| \ll 2\times 10^{-9}, \label{consteq} \ee
where $\dmeV=\dm/\eV^2$.
$F_+(t_{\rm res})\le1$ is also the condition that the $\nu_{\rm s}$ do not come into equilibrium.
Our result here is in good agreement with previous constraints \cite{SSF93a}.

The quasi-static approximation is of course not exact, actually the variables will
always be a bit behind their quasi-static values. This delay becomes
important when the quasi-static value of a variable passes zero, or when the rate of change of
the quasi-static value becomes of
order of the damping rate of the variable.
At resonance, i.e.~when $(\bc-b_T)=0$, we have $\alm_\st=0$, $\bav_\st=0$, $(P_z^-)_\st=0$. Furthermore,
all the quasi-static values change relatively fast near the resonance.

Here is a qualitative discussion of the behavior of the variables close to resonance.
The term that is responsible for the resonance, $\Delta b\equiv\bc-b_T$, only enters into the differential
equation of $\bav$. When $\Delta b$ changes sign, $\bav_\st$ does too. Therefore,
after a short delay, $\bav$ will also change its sign. The other variables do not depend directly
on $\Delta b$; they only feel its change via $\bav$. Thus, they are not sensitive to the time delay
between $\bav$ and $\bav_\st$. Therefore, setting $\bav=\bav_\st$ will merely shift the events by
a negligible amount of time.

Close to resonance, we have $\sbav\approx0$ and $\cbav\approx1$.
Therefore, $\alp$ becomes maximal close to the resonance,
i.e.~$\alp\approx \bs/D$, while $\alm$ becomes minimal so that we can neglect
it in the equations.
$\bwi$ will stay close to its quasi-static value since $\dPz$ does not change significantly:
$\dPz$ freezes out when the rate of change in $(\dPz)_\st$
becomes larger than the damping rate $\kappa$, i.e.~$r_{\dPz}>\kappa$, where $r_{\dPz}$ is given
in (\ref{appdpz}). Since 
$\Delta b\equiv\bc-b_T\ll D$ and $\varepsilon\ll1$ at freeze-out, we get
\be r_{\dPz}\approx -6 H \frac{b_c}{\Delta b}, \quad \kappa\approx \frac{2P_z^+ b_s^2 b_A\Delta b}{D^3}. \ee
Using $T=\TC$, we find that $\Delta b_\fr = -8.1\times 10^{-9} s^{-1} |\dmeV|^{-1/12}c^{11/12} b_T$,
and thus $\varepsilon_\fr = 7.2\times 10^{-3} s |\dmeV|^{5/12} c^{-7/12} \ll 1$, so that
 $(\dPz)_\fr \approx \varepsilon_\fr \Pc$ is indeed very small.

In summary, at resonance we can still approximate the variables $\alp$, $\bav$ and $\bwi$ by their
static values, while we can neglect $\alm$. $\dPz$ is given by its freeze-out
value.

\section{Evolution at Resonance}

We now come to the most interesting feature of the system: $\sin\bav$ changes its sign and becomes
positive. Therefore, the back-reaction of $\bwi$ on $\dPz$,
which is given by the coefficients of the first terms in (\ref{pzm2}) and (\ref{bwi2}),
$b_A \cdot \bs P_z^+ \alp \sin\bav$, also changes its sign to positive.
So while $\dPz$
until now was washed out through this back-reaction, the same back-reaction now amplifies $\dPz$.
We will now derive the consequences of this effect. To this end, we will
assume that the variables $\alfa^\pm$ and $\beda^\pm$ will be given approximately by their
static values. As time passes, these approximations will break down and we will need to find
other approximations.

\subsection{Static Approximation after Resonance}

The change of $\dPz$ becomes interesting again when $r_{\dPz}\le\kappa$ after the resonance,
i.e.~when $\Delta b_0=-\Delta b_\fr$. We will denote the time when this happens with $t_0$.
So until then, $\dPz(t_0)\approx(\dPz)_\fr=\varepsilon_\fr \Pc$. Now (\ref{dpzsim}) has changed to
\be \ddt \dPz = +\kappa (\dPz - \varepsilon P_z^-) \approx +\kappa(\dPz+\varepsilon\Pc).\ee
If we neglect the second term, we get the solution
\be \dPz(t) = \dPz(t_0) \exp\l(\int\limits_{t_0}^t\kappa\,dt\r). \label{sol1}\ee
To solve the integral, we substitute $t$ with $\xi=1-x\ll 1$, where $x=T(t)/\TC$. Then
\be \Delta b = \bc -b_T =\bc^\res x^{-1}-b_T^\res x^5= b_T^\res x^{-1}(1-x^6)\approx
6 b_T^\res \xi, \label{dbapprox}\ee
and
\be d\xi=-\frac{1}{\TC}dT=\frac{5.5\TC^2x^3}{\mpl}dt\approx\frac{5.5\TC^2}{\mpl}dt, \label{dxidt}\ee
where we have expanded $\xi$ and have only taken into account the leading order.
The integral becomes
\ba \int\limits_{t_0}^t\kappa\,dt &=&\frac{\mpl}{5.5 \TC^2} \int\limits_{\xi_0}^{\xi} \kappa d\xi \approx
\frac{\mpl}{5.5 \TC^2}\int\limits_{\xi_0}^{\xi} \frac{2P_z^+\bs^2b_A6b_T^\res\xi}{D^3} d\xi
\nonumber \\
&=& \l.\frac{2P_z^+\bs^2b_A6b_T\mpl}{D^35.5 T^2}\r|_{T=\TC}\int\limits_{\xi_0}^{\xi}\xi d\xi 
\nonumber \\
&=& 5.5 \times 10^{17} s^2 |\dmeV|^{1/6} c^{-11/6}\cdot {\textstyle\ot} (\xi^2-\xi_0^2), \label{int1}\ea
where we have assumed that $\Delta b\ll D$ and have again expanded $\xi$ to leading order.
Furthermore, $\xi_0=\Delta b_0/6b_T^\res=1.35\times10^{-9}s^{-1} |\dmeV|^{-1/12}c^{11/12}$.

The solution (\ref{sol1}) is only valid as long as $\bwi$ and $\alm$ can follow their static
approximations, which are given by (\ref{bwi-st}) and (\ref{alm-st}).
We see that as long as $\Delta b=\bc-b_T\ll D$, the changes in $\bwi_\st$ and $\alm_\st$
will mainly be due to $\dPz$, i.e.~$\dot\bwi_\st/\bwi_\st\approx\dot\alp_\st/\alp_\st\approx
\frac{d}{dt}\dPz/\dPz$. Then the validity of (\ref{sol1}) breaks down
when $D=\ddt\dPz/\dPz=\kappa$, which happens when
$\xi=\xi_1=1.1\times 10^{-14} s^{-2} |\dmeV|^{1/3} c^{7/3}$. We see that the above solution already
breaks down very early, for a large parameter range we even have $\xi_1<\xi_0$. To give
an estimate of how big the integral (\ref{int1}) will be, we set $\xi_0=0$ and get
$\int_{t_\res}^{t_1} \kappa\,dt = 0.8 \times 10^{-11}s^{-2}|\dmeV|^{5/6} c^{17/6}$, which
means that the change in $\dPz$ between $t_\res$ and $t_1$ is negligible.
So we can immediately give up the static approximations for $\bwi$ and $\alm$.

\subsection{Mathematical Pendulum}
\label{pendsec}

After the quasi-static approximation for $\bwi$ breaks down, the first term in (\ref{bwi2})
will be the dominant one due to the rapid growth of $\dPz$, so we neglect the other terms.
In (\ref{pzm2}), we can neglect the terms with the slowly varying $\alm$.
Since $b_A\gg\bs\alp\sbav$, $P_z^-$ will have a much larger effect on $\dot\bwi$ than
on $d(\dPz)/dt$, so that soon $\bwi\gg P_z^-$, and we can thus neglect the third term in
(\ref{pzm2}). Therefore we get the simplified equations
\ba \ddt\dPz&\approx& \bs (\alp\sin\bav)\sin\bwi \\
 \ddt\bwi&\approx& b_A \dPz. \ea
Together, they give the second order equation
\be \ddot\bwi\approx g \sin\bwi, \label{ddotbwi} \ee
where $g=b_A \bs(\alp\sin\bav)$.
We have now found a very simple description of the system.
In the next two subsections, we will discuss the features of this equation.
Afterwards, we discuss the consequences for our system.

We take $t=t_0$ to be the initial time. Then the initial conditions are given
by \be \bwi_0=\bwi_\st\approx\frac{b_A}{D}(\dPz)_\fr\ll 1,\quad \dot\bwi\approx 0. \label{inicond} \ee
For definiteness, we take $\bwi_0$ to be positive, which implies $\dPz>0$.

\begin{figure}
\unitlength1mm
\begin{picture}(121,80)
  \put(25,10){\psfig{file=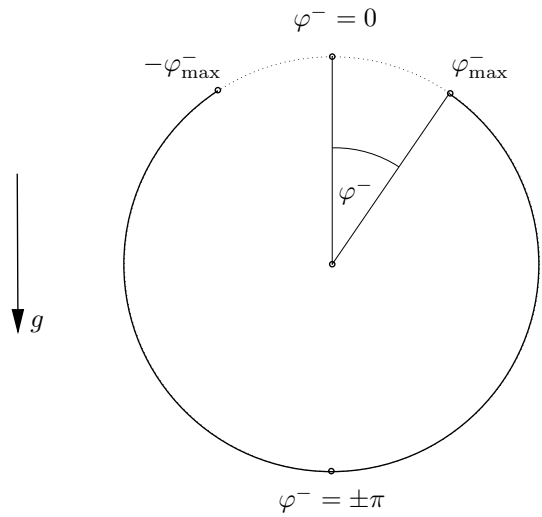,width=7cm}} 
  \put(60,6){\footnotesize $\bwi=\pm\pi$}
  \put(62,70){\footnotesize $\bwi=0$}
  \put(42,64){\footnotesize $-\bwi_{\rm max}$}
  \put(83,64){\footnotesize $\bwi_{\rm max}$}
  \put(68,47){\footnotesize $\bwi$}
  \put(27,30){\footnotesize $g$}
\end{picture}
  \caption{\label{pend-pict}Pictorial view of our pendulum.}
\end{figure}

\begin{figure}
\unitlength1mm
\begin{picture}(121,75)
  \put(15,0){\psfig{file=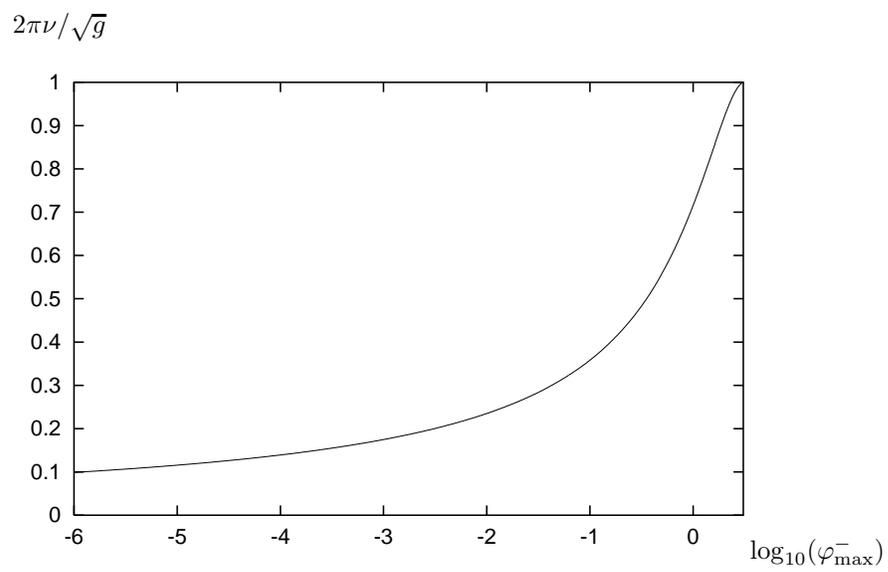,width=10.4cm}} 
  \put(115,0){\footnotesize $\log_{10}(\bwi_{\rm max})$}
  \put(17,70){\footnotesize $2\pi\nu/\sqrt{g}$}
\end{picture}
\begin{center}\parbox{9cm}{
\caption{\label{pend-frec} Frequency of the pendulum as a function of the amplitude.}
}\end{center}
\end{figure}

Let us first assume that $g$ is constant. Then the differential equation corresponds
to a mathematical pendulum, where $g$ is the acceleration of gravity, see Fig.~\ref{pend-pict}.
Usually, we could apply the small angle expansion around the stable minimum of
the potential energy. However, in our case
$\bwi=0$ corresponds to the meta-stable maximum of the potential energy. So our
system will perform large-amplitude oscillations. We denote the amplitude
of the oscillations with $\bwi_{\rm max}$, so small $\bwi_{\rm max}$ corresponds to large amplitude.
For illustration, we can define the analogy of a potential energy $E_\pot=-g(1-\cos\bwi)$
and a kinetic energy $E_\kin=\ot(\dot\bwi)^2$. Conservation of energy
implies that the total energy $E_\tot=E_\pot(t_0)+E_\kin(t_0)={\rm const}$. Since
$E_\tot(t_0)<0$, we know that the system will oscillate around the stable point
$\bwi=\pi$ with constant amplitude $\bwi_{\rm max}=|\bwi_0|$, and that $\bwi$ will never
pass the meta-stable point $\bwi=0$. We can also state that due to the non-linearity of
$\sin\bwi$, the oscillation frequency
of the system, $\nu$, will be smaller than the oscillation frequency $\sqrt{g}/2\pi$
for the linear small-angle approximation.
In fact, for $E_\tot\ra0$, the oscillation frequency goes to zero.
In Fig.~\ref{pend-frec},
we have plotted the ratio $2\pi\nu/\sqrt{g}$ as a function of the amplitude $\bwi_{\rm max}$.

\begin{figure}
\unitlength1mm
\begin{picture}(121,75)
  \put(15,0){\psfig{file=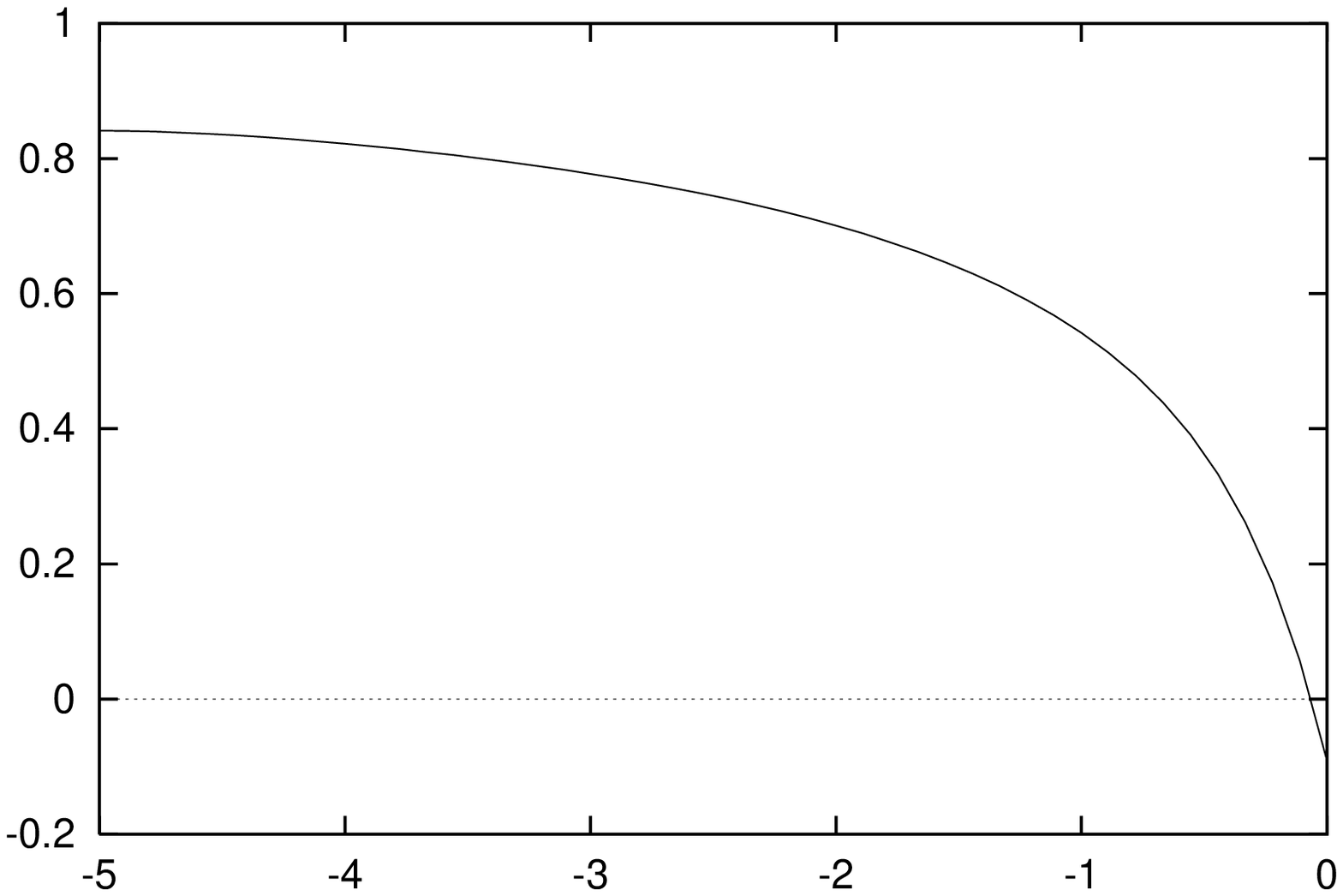,width=10.4cm}} 
  \put(117,0){\footnotesize $\log_{10}(\bwi_{\rm max})$}
  \put(18,68){\footnotesize $\l<\cbwi\r>$}
\end{picture}
\begin{center}\parbox{9cm}{
  \caption{\label{pend-avcbwi}The average $\cbwi$ as a function of the amplitude.}
}\end{center}
\end{figure}

Another important fact is that the time average of $\cbwi$, $\l<\cbwi\r>$, is greater than 0 for
small $\bwi_{\rm max}$, i.e.~for large amplitudes. This results from the fact that the system
develops relatively slowly close to the turning points $\pm\bwi_{\rm max}$, where the
kinetic energy is small. We have plotted $\l<\cbwi\r>$ as a function of
the amplitude $\bwi_{\rm max}$ in Fig.~\ref{pend-avcbwi}.

Next, we discuss the differential equation with time dependent $g$. It is sufficient
if we deduce the effect on the amplitude $\bwi_{\rm max}$; we can then use $\bwi_{\rm max}$ and
Figs.~\ref{pend-frec} and \ref{pend-avcbwi} to
derive the oscillation frequency and $\l<\cbwi\r>$.

We will assume that $\dot g$ is approximately constant over the time of half the oscillation period,
$\tau/2$, where $\tau=1/\nu\ge2\pi/\sqrt{g}$.
Then it is easy to calculate the change in the amplitude $d\bwi$ during the time $dt=\tau/2$.
Taking the differential of the potential energy, we get
\be dE_\pot^{\rm max}=-dg~ (1-\cbwi_{\rm max}) + g~ d(\cbwi) \label{depot}\ee
Since the acceleration of gravity, $g$, is no longer constant, the total energy is not conserved,
so that
\ba \dot E_\tot &=& \dot E_\kin + \dot E_\pot = \dot\bwi\ddot\bwi-g\sbwi\dot\bwi-\dot g(1-\cbwi) \nonumber\\
&=& -\dot g(1-\cbwi), \ea
where the first two terms cancel due to the differential equation (\ref{ddotbwi}).
We get the differential
\be dE_\tot\approx - dg~ \l(1-\l<\cbwi\r>\r), \label{detot}\ee
where we have time averaged $\cbwi$.
Equating (\ref{depot}) and (\ref{detot}), we get
\be d(\cbwi_{\rm max}) = -\frac{dg}{g} \l(\cbwi_{\rm max} - \l<\cbwi\r>\r), \ee
If we use $dg = \dot g \,dt=\dot g\, \tau/2$ and $d(\cos|\bwi|)=d(\cos|\bwi|)=-\sin|\bwi|d|\bwi|$, we finally get
\be d|\bwi_{\rm max}|=-\frac{\dot g \tau/2}{g \sin|\bwi_{\rm max}|}\l(\cbwi_{\rm max} - \l<\cbwi\r>\r) \ee
as the change in the amplitude during one half oscillation,
where of course $\cbwi_{\rm max} \ge \l<\cbwi\r>$. We see that $\dot g>0$ will result in a decreasing amplitude,
i.e.~increasing $|\bwi_{\rm max}|$. Analogously, $\dot g<0$ will result in an increasing amplitude.
If thereby $|\bwi_{\rm max}|+d|\bwi_{\rm max}|<0$, $E_\tot>0$, so the system passes the meta-stable point and
accelerates on the other side instead of turning.
Then it is better to use (\ref{detot}) instead of the differential for the amplitude to describe
how the system evolves.

Our approximation breaks down if $|\bwi_{\rm max}|+d|\bwi_{\rm max}|\ll1$,
since then $(\cbwi_{\rm max} - \l<\cbwi\r>)\ra 0$,
$\sin^{-1}\bwi_{\rm max}\ra \infty$ and $\tau\ra \infty$.
Also, $\dot g ={\rm const}$ might not apply any longer.

Now we can already make a statement about the neutrino asymmetry, represented by $\dPz=\dot\bwi/b_A$.
If $E_\tot<0$, the asymmetry will change sign at the turning points $\pm\bwi_{\rm max}$, i.e.~the
asymmetry oscillates. If $E_\tot>0$, the asymmetry will oscillate
between $\sqrt{2E_\tot}/b_A$ and $\sqrt{2(E_\tot+g)}/b_A$,
but will not change sign.

\subsection{The Factor {\boldmath$g$}}

The next step is to derive $\dot g$.
The factors $b_A$ and $\bs$ in $g$ only vary slowly with time, and we will
take them to be constant. Thus the main contribution to $\dot g$ comes from $\alp\sin\bav$.
For convenience, we define two new variables, $\alpha=\alp\sbav$ and $\beta=\alp\cbav$. Then
$g=\bs b_A\alpha$, so that $\dot g\propto \dot \alpha$. We get
\ba \dot\alpha&=& -D\alpha + (\bc-b_T)\beta \label{aeq}\\
    \dot\beta &=& -D\beta  - (\bc-b_T)\alpha + \bs \cos\bwi, \label{beq}\ea
where we have neglected all terms dependent on $\alm$.

Initially, $\bwi\ll1$, so we will first consider the simpler case where $\cbwi=1$, $\sbwi=0$.
Then we can apply the quasi-static approximation for $\alp$ and $\bav$. Thus,
\ba \alpha_\st &=& \frac{\bs(\bc-b_T)}{D^2+(\bc-b_T)^2} \label{alpha-st-pl} \\
    \beta_\st  &=& \frac{\bs D}{D^2+(\bc-b_T)^2}. \ea
Now we can determine $\dot g=\bs b_A\dot\alpha_\st$ with the help of
\be \dot \alpha_\st = H~\frac{\bs(-b_T\Delta b^2 + D^2 \Delta b + D^2 \bc)}{[D^2+\Delta b^2]^2}, \ee
where $\Delta b=\bc-b_T$.

For small $\Delta b$, $\dot\alpha_\st$ is positive.
However, $\dot\alpha$ changes sign when the numerator becomes zero, i.e.
\be \Delta b_\alpha = \frac{D^2}{2b_T}\l(1\mp\sqrt{1+4\frac{\bc b_T}{D^2}}\r). \ee
Using $\bc\approx b_T=D/k_D$, we find that the term $\bc b_T/D^2\approx k_D^{-2}\gg1$. Therefore,
\be \Delta b_\alpha \approx D = k_D b_T. \ee
This result can be expressed in terms of the
dimensionless variable $\xi=1-T(t)/T$. Then we get $\xi_\alpha\approx k_D/6$, see (\ref{dbapprox}).

\begin{figure}
\unitlength1mm
\begin{picture}(121,76)
  \put(15,0){\psfig{file=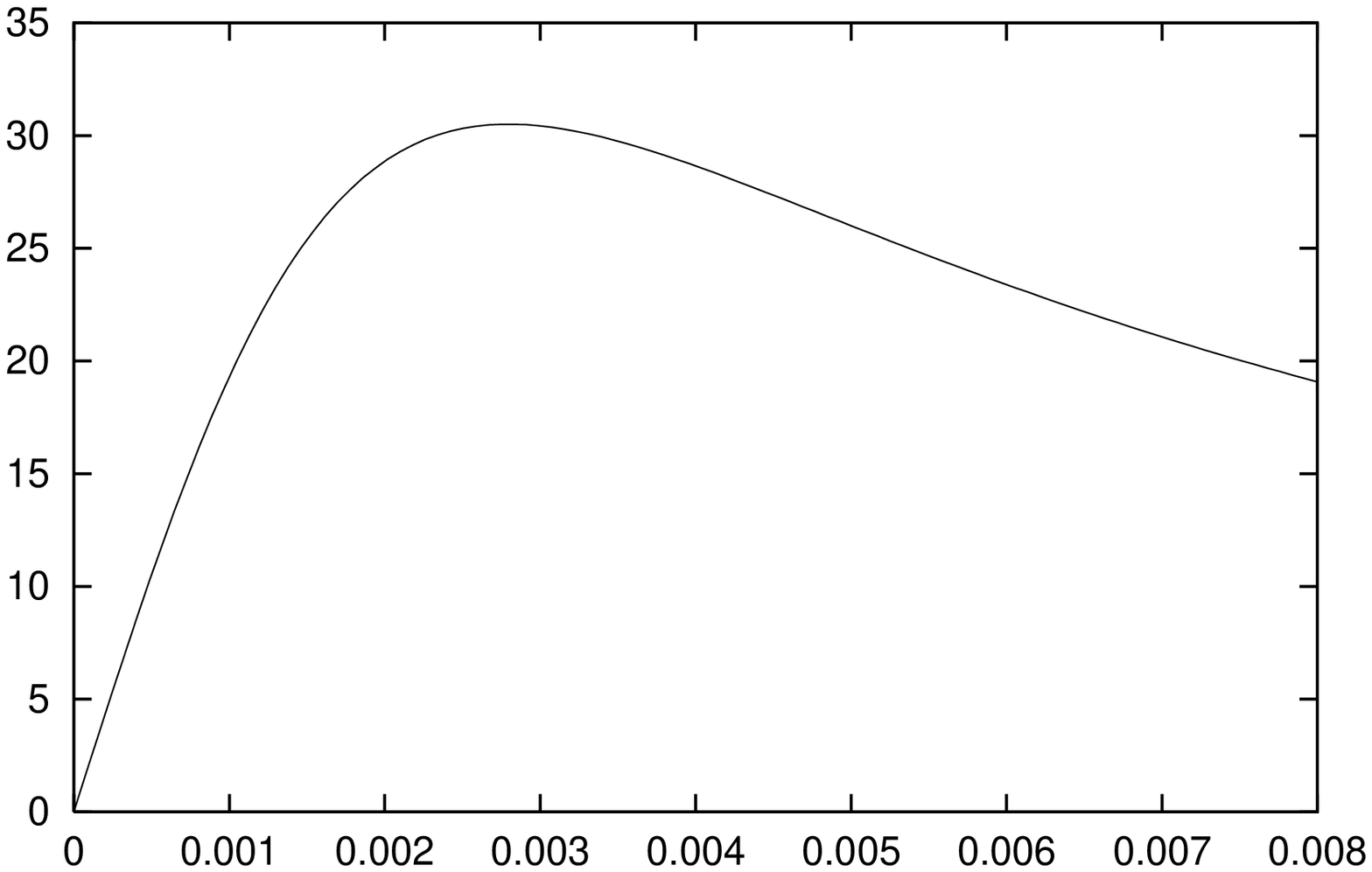,width=10.4cm}} 
  \put(120,0){\footnotesize $\xi$}
  \put(18,68){\footnotesize $\frac{c}{s}\alpha_\st$}
\end{picture}
\begin{center}\parbox{9cm}{
\caption{\label{alpha-xi} Evolution of $\frac{c}{s}\alpha_\st$, given in
(\ref{alpha-st-pl}),
 as a function of $\xi=1-T/\TC$.}
}\end{center}
\unitlength1mm
\begin{picture}(121,76)
  \put(15,0){\psfig{file=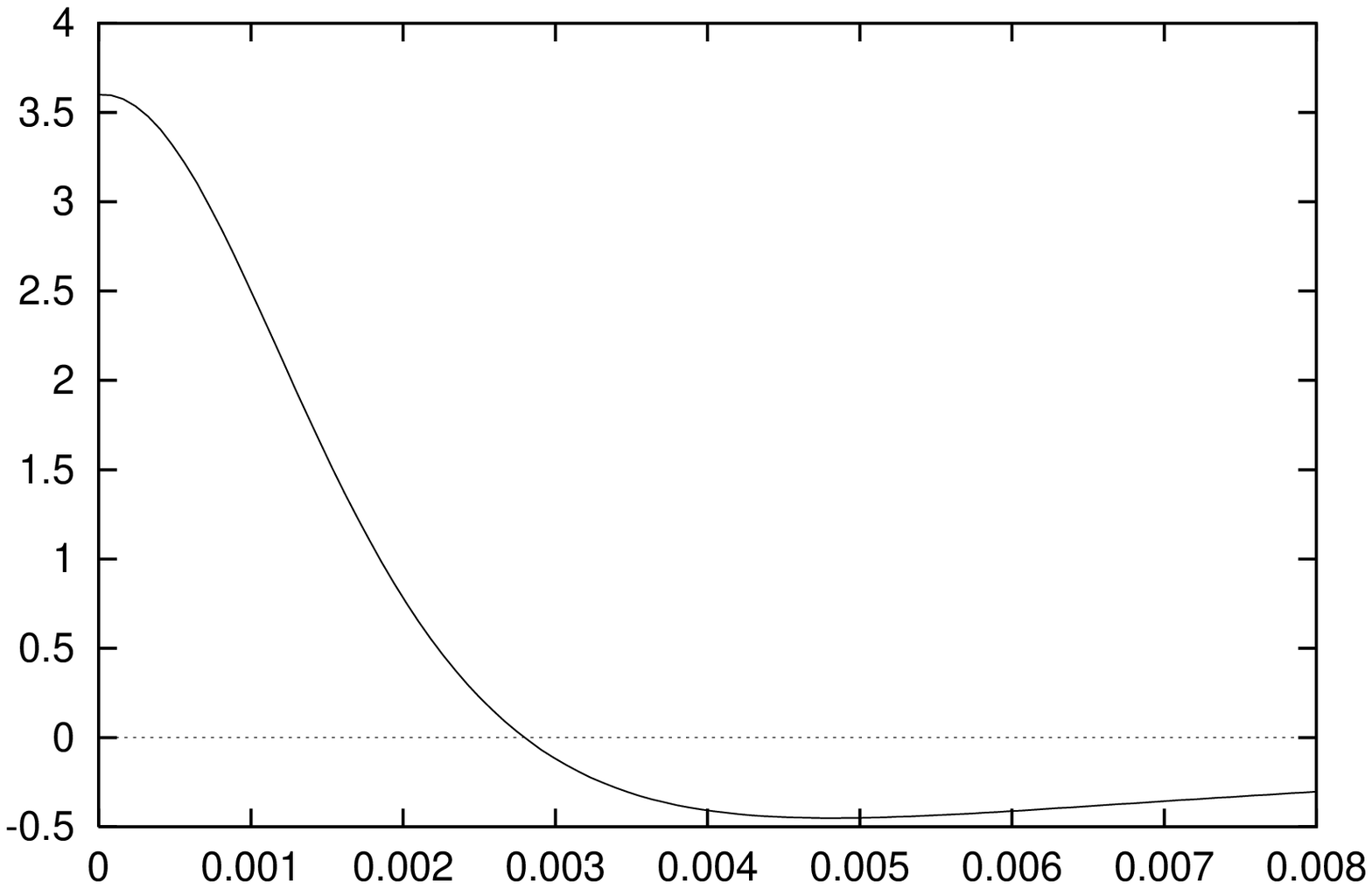,width=10.4cm}} 
  \put(120,0){\footnotesize $\xi$}
  \put(10,68){\footnotesize $\frac{c}{s}\frac{d}{d\xi}\alpha_\st~[10^{+3}]$}
\end{picture}
\begin{center}\parbox{9cm}{\caption{\label{dotalpha-xi} Evolution of $\frac{c}{s}
\frac{d}{d\xi}\alpha_\st = \frac{c}{s}\frac{\TC}{HT}\,\dot\alpha_\st$
as a function of $\xi$.}}\end{center}
\end{figure}

In summary, $g$ will increase at first and then decrease for $\Delta b>\Delta b_\alpha$.
We have plotted the parameter-independent dimensionless variables $\alpha_\st c/s$ and
$\frac{d}{d\xi}\alpha_\st c/s= \dot\alpha_\st \TC c/HTs$
as a function of $\xi$ in Figs.~\ref{alpha-xi} and \ref{dotalpha-xi}, respectively.
We can see that $\dot\alpha_\st$ is constant to lowest order for very small $\xi$.
Therefore, we expand $g$ in terms of $\xi$,
\be g = 6\l.\frac{b_A\bs^2b_T}{D^2}\r|_{\TC} \xi+\mathcal{O}(\xi^2). \label{glin}\ee
This approximation is valid for $\xi<\xi_\alpha$.

Now we consider the situation where $\bwi$ changes in time.
We assume that the oscillation period $\tau=\mathcal{O}(1/\sqrt{g})$ is much smaller than the
damping time scales $1/D$ and $1/\Delta b$. Then we can approximate $\sbwi$ and $\cbwi$ by their
time average values, $0$ and $\l<\cbwi\r>$, respectively, when describing the evolution of
$\alpha$ and $\beta$. Thus
\ba \dot\alpha&\approx& -D\alpha + \Delta b \beta \label{aeq2}\\
    \dot\beta &\approx& -D\beta  - \Delta b \alpha + \bs \l<\cos\bwi\r>. \label{beq2}\ea
Note that here we can easily neglect $\alm$, since $\alm_\st\propto\l<\sbwi\r>=0$.
The coefficient $\Delta b$ in (\ref{aeq2}) and (\ref{beq2}) induces an oscillation
of the two variables, which will be damped by the coefficient $D$
toward their quasi-static approximations, given by
\ba \alpha_\st &=& \frac{\bs\Delta b}{D^2+\Delta b^2} \l<\cos\bwi\r>, \\
    \beta_\st  &=& \frac{\bs D}{D^2+\Delta b^2} \l<\cos\bwi\r>. \ea

\subsection{Proof for Oscillations}

It is now easy to describe the evolution of our system.
We will restrict ourselves to proving that the pendulum, and therefore $\dPz$, will oscillate for
a certain range of mixing parameters. $\dPz$ will change sign at least once
if $\dot g$ is positive during the first oscillation of $\bwi$, since then the amplitude $\bwi_{\rm max}$
decreases.

We start with the initial values given in (\ref{inicond}). For small $\bwi$, we can linearize
the differential equation (\ref{ddotbwi}), so that
\be \ddot\bwi\approx g \bwi. \label{ddotbwi2} \ee
Furthermore, we linearize $g$ as we have done in (\ref{glin}) and substitute 
$dt$ by $d\xi$, as shown in (\ref{dxidt}). Then we get
\be \frac{d^2 \bwi}{d\xi^2} = h \xi\bwi, \ee
where $h=1.02\times 10^{20} s^2 c^{-4/3} |\dmeV|^{2/3}$.
The solutions for this equation are the Airy-functions.

When $\bwi=\mathcal{O}(1)$, the linear approximation breaks down, and $\bwi$ will start
oscillating with a frequency $\nu\simlt\sqrt{g}/2\pi$.
We want $\dot g\propto\dot\alpha$ to be positive, so this first oscillation has to
start before $\xi_\alpha$. Thus we demand $\xi_{\bwi}<\xi_\alpha$, where $\xi_{\bwi}$ is defined by
$\bwi(\xi_{\bwi})=1$. We find numerically that we can fit the solution of this condition with
\be s^{3.1}|\dmeV| \ge 10^{-15.1} \label{lasti}\ee
for $|\dm|=10^{-9}\mbox{--}10^5 \eV^2$ with an error smaller than 3\%.

Of course, we need to take into account that the quasi-static solution 
for $\alpha$ given in \ref{alpha-st-pl} breaks down, since $\l<\bwi\r><1$ during oscillations.
However, as we can see in (\ref{aeq2}) and (\ref{beq2}), $\dot\alpha$
does not depend on $\l<\cbwi\r>$ directly, but indirectly through
$\Delta b \beta$. Thus, the change in $\l<\cbwi\r>$ will first affect $\dot\alpha$
after a time scale $1/\Delta b$. So if the condition $\Delta b\ll\sqrt{g}$ holds at $\xi=\xi_{\bwi}$,
then $\dot\alpha$ stays positive during the first few oscillation periods, and thus the
amplitude will decrease.

Since $\Delta b/\sqrt{g}\propto\sqrt{\xi}$, we can use the stronger condition
$\Delta b/\sqrt{g}\ll1$ for $\xi=\xi_\alpha$. Then we get
\be s|\dmeV|^{-1/6}\gg3\times 10^{-6}. \label{lastcondo} \ee
We have now found the conditions for which $\dPz$ oscillates.

We can also estimate
the amplitude with which the neutrino asymmetry $\dPz$ oscillates.
An upper bound is given if we use the maximal value for $g$, i.e.~at $\xi=\xi_\alpha$,
and assume $E_\tot=0$. Then $\dPz=\dot\bwi/b_A$ is maximal at $\bwi=\pi$, i.e. $E_\kin=-E_\pot=2g$.
We derive
\ba (\dPz)_{\rm max} &=& \frac{1}{b_A}
   (\dot\bwi)_{\rm max} =
  \frac{1}{b_A} \sqrt{2(E_{\rm kin})_{\rm max}} \nonumber\\
   &=& \frac{1}{b_A} \sqrt{4g(\xi_\alpha)}\approx
  \frac{2}{b_A}\sqrt{\frac{\bs^2b_A}{2D}} \nonumber\\
  &\approx& 1.4\times 10^{-2}s|\dmeV|^{1/6}c^{-5/6}.\ea
We can compare this result with the numerical solution given in Section \ref{numsolsec}.
Numerically we find during the period where $\dPz$ oscillates that $\dPz\le7.53\times10^{-7}$
for the parameters $\dm=-1\eV^2$, $s=10^{-4}$.
This is in very good agreement with our analytical value $(\dPz)_{\rm max}=
1.4\times10^{-6}$. The difference is due to the fact that when $g$ is maximal, the amplitude
of $\bwi$ is not maximal, and thus a factor of $\sqrt{2}$ has to be replaced by $\sqrt{1+\cos\bwi_{\rm max}}$.
Furthermore, $\alpha$ is slightly decreased due to $\l<\cbwi\r><1$.

\chapter{Conclusions}

\begin{figure}
\unitlength1mm
\begin{picture}(121,130)
  \put(10,15){\psfig{file=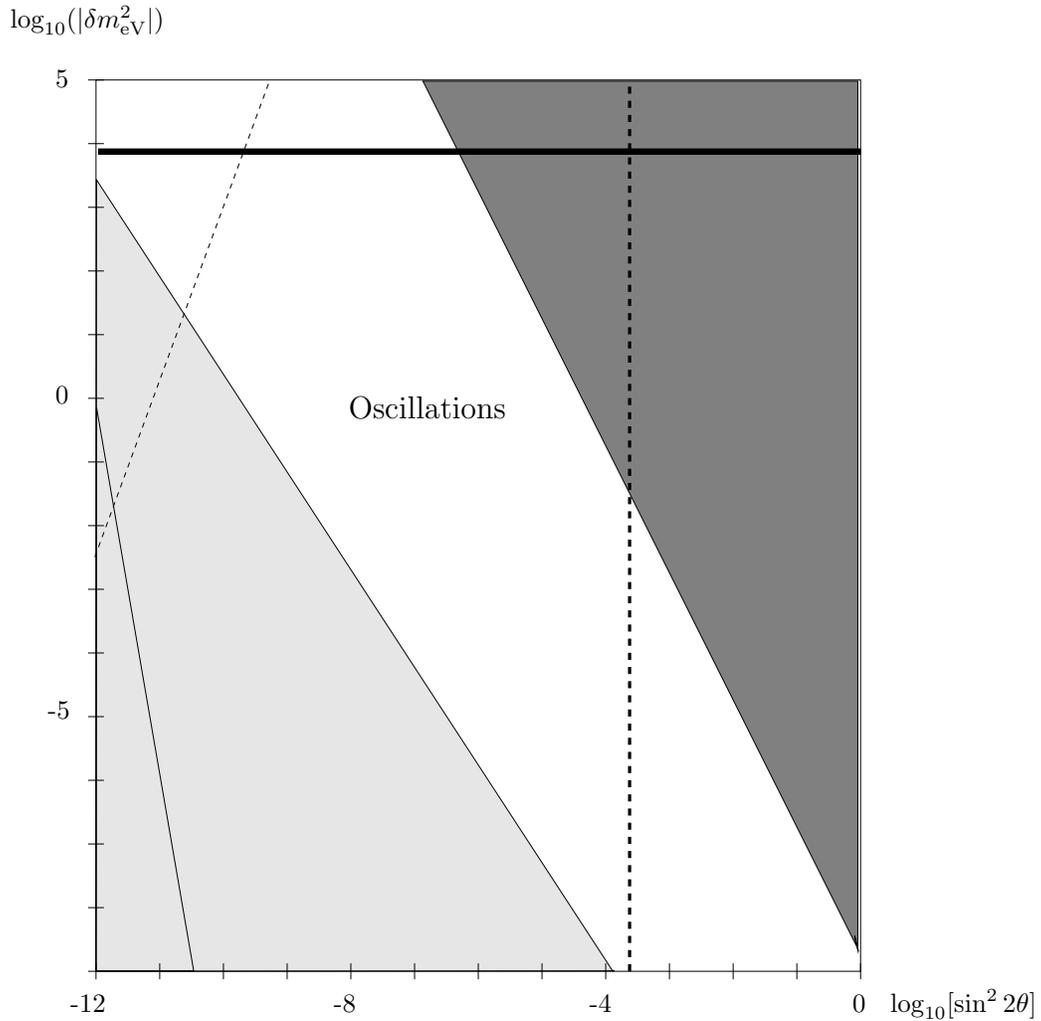,width=10.4cm}} 
  \put(117,11){\footnotesize $\log_{10}[\sin^22\theta]$}
  \put(0,142){\footnotesize $\log_{10}(|\dmeV|)$}
  \put(45,90){Oscillations}
  \put(112,11){\footnotesize 0}
  \put(77,11){\footnotesize -4}
  \put(43,11){\footnotesize -8}
  \put(8,11){\footnotesize -12}
  \put(6,134){\footnotesize 5}
  \put(6,92){\footnotesize 0}
  \put(5,50){\footnotesize -5}

\end{picture}
  \caption{\label{range} Bounds on the parameter space derived in our work. 
In the central white region, our proof of asymmetry oscillations is valid.
In the dark-shaded region, the sterile neutrinos come into equilibrium, see (\ref{consteq}).
In the light-shaded region, the factor $g$ starts decreasing again before
$\bwi=\mathcal{O}(1)$, see (\ref{lasti}); we cannot say whether the asymmetry oscillates in this
region of parameter space. Below the lower left line, $\dPz$ does not reach its quasi-static
solution before resonance, see (\ref{dpzqs}). Our small-angle approximation, $s\ll k_D$, is
valid only to the left of the thick dashed line. To the left of the thin dashed line,
the condition (\ref{lastcondo}) does not apply; still, there might be oscillations.
Above the horizontal thick line, the universe would be over-closed \cite{KTTEU}.}
\end{figure}

We have analytically examined neutrino oscillations between
a sterile and an active neutrino in the Early Universe, neglecting the neutrino
momentum distribution. Our main achievement has been to prove
that this system, which is known to create
a large neutrino asymmetry, exhibits oscillations of this neutrino
asymmetry for a large range of mixing parameters, as shown in
Fig.~\ref{range}. We conclude that these
asymmetry oscillations, which have been encountered in only some
of the numerical calculations, can not arise from numerical instabilities.

With the methods presented in our work,
the complete analytical treatment of this system seems to have become feasible.
Naturally, the next step will be to include the effects from the neutrino
momentum distribution on the oscillatory behavior.
Then the analytical approach will clearly be superior to the numerical one, which continues to
encounter many difficulties \cite{BF99a}. Finally, the duration of the
asymmetry oscillations and the power law of the asymmetry growth after
the oscillations have ceased should be derived.

Understanding the mechanism of $\nu_\alpha$--$\nu_{\rm s}$
oscillations in the Early Universe is very important.
On the one hand, if the primordial abundances are determined with sufficient
precision, it becomes possible to constrain the mixing parameters of the
$\nu_\alpha$--$\nu_{\rm s}$ system. As a result, some of the models which have been
proposed to explain the current experimental situation could be excluded.

On the other hand, if future neutrino experiments, such as
MiniBooNE \cite{MINIBOONE}, prove the existence of a sterile neutrino,
it is necessary to understand the impact of the $\nu_{\rm s}$
on the primordial abundances to test whether BBN is a consistent theory.

\begin{appendix}

\chapter{Quasi-Static Approximation and its Range of Validity}
\label{AppSTAT}

The quasi-static approximation of a variable $x$ which follows the differential
equation
\be \frac{\partial}{\partial t} x(t) = -D(t)x(t)+f(t) \ee
is given by
\be x_\st=\frac{f}{D}. \ee
If $f$ and $D$ depend on time, $x_\st$ also changes its value. Therefore,
$x_\st$ will only be a good approximation if the quasi-static condition is
fulfilled, i.e.
\be |\dot x_\st| < D x_\st. \ee
We have derived all values $r_x\equiv\dot x_\st/x_\st$ for our system of differential
equations (\ref{pzp2})--(\ref{bwi2}) assuming that all variables except for
$P_z^+$ are at their static values, where $H=-\dot T/T$.
\ba r_{\alp}&\equiv&\frac{\dot \alp_\st}{\alp_\st}
        = -6 H \l(-1 + \frac{\bc(\bc-b_T)}{D^2+(\bc-b_T)^2}\r), \\
    r_{\bav}&\equiv&\frac{\dot \bav_\st}{\sin \bav_\st}
        = -6 H \frac{-\bc}{\bc-b_T} \frac{D}{\sqrt{D^2+(\bc-b_T)^2}}.\ea
Here we have divided by $\sin \bav_\st$ instead of $\bav_\st$, since the
differential equation for $\bav$ has the form  $\frac{\partial}{\partial t} x(t) = -D(t)\sin[x(t)]+f(t)$.
In the case of $\bwi$, we can neglect the $\sin$ as long as $\bwi\ll1$. Furthermore,
\ba r_{P_z^-}&\equiv&\frac{(\dot P_z^-)_\st}{(P_z^-)_\st}
        = - \frac{\dot \varepsilon}{\varepsilon} ~ \frac{\varepsilon}{1+\varepsilon}, \\
    r_{\dPz}&\equiv&\frac{\frac{d}{dt}(\dPz)_\st}{(\dPz)_\st}
        =   \frac{\dot \varepsilon}{\varepsilon} ~ \frac{1}{1+\varepsilon}, \label{appdpz}\\
    r_{\alm}&\equiv&\frac{\dot \alm_\st}{\alm_\st}= r_{\alp} + r_{P_z^-} \nonumber\\
        &=& - H \l(-8 -6 \frac{\bc}{\bc-b_T} \l(1-3\frac{(\bc-b_T)^2}{D^2+(\bc-b_T)^2}\r)\r) + 
              r_{\dPz}, \\
    r_{\bwi}&\equiv&\frac{\dot \bwi_\st}{\bwi_\st}= -6 H \frac{\bc}{\bc-b_T} + r_{P_z^-}\nonumber\\
        &=& - H \l(-2+12\frac{\bc(\bc-b_T)}{D^2+(\bc-b_T)^2}\r) + r_{\dPz},  \ea
where
\ba \varepsilon &=& - \frac{D^2+(\bc-b_T)^2}{2P_z^+b_A(\bc-b_T)},  \\
 \frac{\dot \varepsilon}{\varepsilon}
         &=& - H \l(2+6\frac{\bc}{\bc-b_T}\l(1-2\frac{(\bc-b_T)^2}{D^2+(\bc-b_T)^2}\r)\r).  \ea

\end{appendix}


\begin{thebibliography}{99}
\frenchspacing

\bibitem{solarexp}
B.T. Cleveland et al. Nucl. Phys. B (Proc. Suppl.) {\bf 38}, 47 (1995);
K.S. Hirata et al., Phys. Rev. {\bf D44}, 2241 (1991);
GALLEX Collaboration, Phys. Lett. {\bf B388}, 384 (1996);
J.N. Abdurashitov et al., Phys. Rev. Lett. {\bf 77}, 4708 (1996).

\bibitem{Atmos} Y. Fukuda, {\it et al.}, Phys. Rev. Lett. {\bf 81}, 1562 (1998);
Y. Fukuda, {\it et al.}, Phys. Lett. {\bf B433}, 9 (1998);
Y. Fukuda, {\it et al.}, Phys. Lett. {\bf B436}, 33 (1998).

\bibitem{99LSND}
C. Athanassopoulos et al. Phys. Rev. Lett. {\bf 75} (1995) 2650; 
C. Athanassopoulos et al. Phys. Rev. {\bf C58} (1998) 2489.

\bibitem{P57} B. Pontecorvo, Zh. Eksp. Teor. Fiz. {\bf 33}, 549 (1957)/
Sov. Phys. JETP {\bf 7},172 (1958).

\bibitem{raffi99a} With kind permission from G. Raffelt, to be published in {\it Proceedings 1998 Summer School in High-Energy Physics
and Cosmology, ICTP}, hep-ph/9902271.

\bibitem{PDG98a} Particle Data Group, {\it Review of Particle Physics}, European
Physical Journal {\bf C3} (1998), 1.

\bibitem{ed} H. Nunokawa, J.T. Peltoniemi, A. Rossi, J.W.F. Valle,
Phys. Rev. {\bf D56}, 1704 (1997).

\bibitem{KTTEU}  E.W. Kolb, M.S. Turner, The Early Universe, Addison-Wesley (1990).

\bibitem{EKMBD90f} K. Enqvist, K. Kainulainen, J. Maalampi, Phys. Lett. {\bf B244}, 186 (1990);
 Phys. Lett. {\bf B249}, 531 (1990); Nucl. Phys {\bf B349}, 754 (1991);\\
R. Barbieri, A. Dolgov, Phys. Lett. {\bf B237}, 440 (1990).

\bibitem{BD91a}  R. Barbieri, A. Dolgov, Nucl. Phys. {\bf B349}, 743 (1991).

\bibitem{FTV96a} R. Foot, M.J. Thomson, R.R. Volkas, Phys. Rev. {\bf D53}, 5349 (1996).

\bibitem{FV95a}  R. Foot, R.R. Volkas, Phys. Rev. Lett. {\bf 75}, 4350 (1995).

\bibitem{FV96a}  R. Foot, R.R. Volkas, Phys. Rev. {\bf D55}, 5147 (1997).

\bibitem{SF98a}  X. Shi, G.M. Fuller, Phys. Rev. {\bf D59}, 063006 (1999).

\bibitem{F98a}   R. Foot, Astropart. Phys. {\bf 10}, 253 (1999).

\bibitem{FV98a}  R. Foot, R.R. Volkas, astro-ph/9811067.

\bibitem{SF98b}  X. Shi, G.M. Fuller, astro-ph/9812232.

\bibitem{FV97a}  R. Foot, R.R. Volkas, Phys. Rev. {\bf D59}, 029901 (1999).

\bibitem{BFV98a} N.F. Bell, R. Foot, R.R. Volkas, Phys. Rev. {\bf D58}, 105010 (1999).

\bibitem{FV99a}  R. Foot, R.R. Volkas, hep-ph/9904336.

\bibitem{S96a}   X. Shi, Phys. Rev. {\bf D54}, 2753 (1996).

\bibitem{EKS99a} K. Enqvist, K. Kainulainen, A. Sorri, hep-ph/9906452.

\bibitem{SF99a}  X. Shi, G.M. Fuller, Phys. Rev. Lett. (to be published), astro-ph/9904041.

\bibitem{SEMIKOX} A.D. Dolgov, S.H. Hansen, S. Pastor, D.V. Semikoz, hep-ph/9910444.

\bibitem{B99a}   P. Di Bari, hep-ph/9911214.

\bibitem{ASF99a} X. Shi, G.M. Fuller, K. Abazajian, astro-ph/9904052.

\bibitem{BLL99a} P. Di Bari, P. Lipari, M. Lusignoli, hep-ph/9907548.

\bibitem{BVW98a} N.F. Bell, R.R. Volkas, Y.Y.Y. Wong, Phys. Rev. {\bf D59}, 113001 (1999).

\bibitem{ASF99b} X. Shi, G.M. Fuller, K. Abazajian, Phys. Rev. {\bf D60}, 063002 (1999).

\bibitem{F99a}   R. Foot, hep-ph/9906311.

\bibitem{CK99a}  M.V. Chizhov, D.P. Kirilova, hep-ph/9908525.

\bibitem{FV96b}  R. Foot, R.R. Volkas, Astropart. Phys. {\bf 7}, 283 (1997).

\bibitem{ASF99c} X. Shi, G.M. Fuller, K. Abazajian, astro-ph/9908081.

\bibitem{SF98c}  X. Shi, G.M. Fuller, Phys. Rev. Lett. {\bf 82}, 2832 (1999).

\bibitem{ASF99d} K. Abazajian, X. Shi, G.M. Fuller, astro-ph/9909320.

\bibitem{CK99b}  M.V. Chizhov, D.P. Kirilova, hep-ph/9909408.

\bibitem{O91a}   K. A. Olive, D.N. Schramm, D. Thomas, T.P. Walker, Phys. Lett. {\bf B265}, 239 (1991).

\bibitem{FO98a}  B.D. Fields, K.A. Olive, Ap.J. {\bf 506}, 177 (1998).

\bibitem{BT98ab} S. Burles, D. Tytler, Ap.J. {\bf 499}, 699 (1998); \\
                 S. Burles, D. Tytler, Ap.J. {\bf 507}, 732 (1998).

\bibitem{WCLFLVB97} J.K. Webb, R.F. Carswell, K.M. Lanzetta, R. Ferlet, M. Lemoine, A. Vidal-Madjar,
D.V. Bowen, Nature {\bf 388}, 250 (1997); \\
D. Tytler, S. Burles, L. Lu, X.-M. Fan, A. Wolfe, B.D. Savage, astro-ph/9810217, to appear in Astron.J..

\bibitem{OSW99a} K.A. Olive, G. Steigman, T.P. Walker, astro-ph/9905320.

\bibitem{MINIBOONE} A.O. Bazarko, hep-ex/9906003.

\bibitem{KPNPA}  C.W. Kim, A. Pevsner, Neutrinos in Physics and Astrophysics,
Harwood Academic Publishers (1993).

\bibitem{RAFFI}  G.G. Raffelt, Stars as Laboratories for Fundamental Physics, The
University of Chicago Press (1996).

\bibitem{NR88a}  D. N\"otzold, G. Raffelt, Nucl. Phys. {\bf B307}, 924 (1988).

\bibitem{MT94a}  B.H.J. McKellar, M.J. Thomson, Phys Ref. {\bf D49}, 2710 (1994).

\bibitem{STOD87} L. Stodolsky, Phys. Rev. {\bf D36}, 2273 (1887).

\bibitem{SSF93a} X. Shi, D.N. Schramm, B.D. Fields, Phys. Rev. {\bf D48}, 2563 (1993).

\bibitem{BF99a} P. Di Bari, R. Foot, hep-ph/9912215.

\end{thebibliography}
\end{document}